\documentclass[iop]{aastex631}
\usepackage{amsmath}
\usepackage{amssymb}
\usepackage{mathbbol}
\usepackage{dsfont}

\DeclareGraphicsExtensions{.eps,.eps.gz,.epsi}

\shorttitle{S-PLUS: Photometric Re-calibration}
\shortauthors{Xiao et al.}
\begin{document}

\title{S-PLUS: Photometric re-calibration with the Stellar Color Regression Method and an Improved Gaia XP Synthetic Photometry Method}

\author
{Kai Xiao\altaffilmark{1,2}
Yang Huang\altaffilmark{1,3}
Haibo Yuan\altaffilmark{2,4}
Timothy C. Beers\altaffilmark{5}
Bowen Huang\altaffilmark{2,4}
Shuai Xu\altaffilmark{2,4}
Lin Yang\altaffilmark{6}
Felipe Almeida-Fernandes\altaffilmark{7}
H\'elio D. Perottoni\altaffilmark{8}
Guilherme Limberg\altaffilmark{7,9,10}
William Schoenell\altaffilmark{11}
Tiago Ribeiro\altaffilmark{12}
Antonio Kanaan\altaffilmark{13}
Natanael Gomes de Olivira\altaffilmark{14}
}
\altaffiltext{1}{School of Astronomy and Space Science, University of Chinese Academy of Sciences, Beijing 100049, People's Republic of China; huangyang@ucas.ac.cn}
\altaffiltext{2}{Institute for Frontiers in Astronomy and Astrophysics, Beijing Normal University, Beijing, 102206, China; yuanhb@bnu.edu.cn}
\altaffiltext{3}{CAS Key Lab of Optical Astronomy, National Astronomical Observatories, Chinese Academy of Sciences, Beijing 100012, People's Republic of China}
\altaffiltext{4}{Department of Astronomy, Beijing Normal University, Beijing, 100875, People's Republic of China}
\altaffiltext{5}{Department of Physics and Astronomy and JINA Center for the Evolution of the Elements (JINA-CEE), University of Notre Dame, Notre Dame, IN 46556, USA}
\altaffiltext{6}{Department of Cyber Security, Beijing Electronic Science and Technology Institute, Beijing, 100070, China}
\altaffiltext{7}{Universidade de S\~ao Paulo, Instituto de Astronomia, Geof\'isica e Ci\^encias Atmosf\'ericas, Departamento de Astronomia, \\ SP 05508-090, S\~ao Paulo, Brazil}
\altaffiltext{8}{Nicolaus Copernicus Astronomical Center, Polish Academy of Sciences, ul. Bartycka 18, 00-716, Warsaw, Poland}
\altaffiltext{9}{Department of Astronomy \& Astrophysics, University of Chicago, 5640 S. Ellis Avenue, Chicago, IL 60637, USA}
\altaffiltext{10}{Kavli Institute for Cosmological Physics, University of Chicago, 5640 S. Ellis Avenue, Chicago, IL 60637, USA}
\altaffiltext{11}{GMTO Corporation, 465 N. Halstead Street, Suite 250, Pasadena, CA 91107, USA}
\altaffiltext{12}{Rubin Observatory Project Office, 950 N. Cherry Ave., Tucson, AZ 85719, USA}
\altaffiltext{13}{Departamento de F\'isica, Universidade Federal de Santa Catarina, Florian\'opolis, SC 88040-900, Brazil}
\altaffiltext{14}{Valongo Observatory, Federal University of Rio de Janeiro, Rio de Janeiro, 2008090, Brazil}



\begin{abstract}
We present a comprehensive re-calibration of medium- and broad-band photometry from the Southern Photometric Local Universe Survey (S-PLUS) by leveraging two approaches: an improved Gaia XP Synthetic Photometry (XPSP) method with corrected Gaia XP spectra, the Stellar Color Regression (SCR) method with corrected Gaia EDR3 photometric data and spectroscopic data from LAMOST DR7. Through the use of millions of stars as standards per band, we demonstrate the existence of position-dependent systematic errors, up to 23\,mmag for the Main Survey region, in the S-PLUS DR4 photometric data. A comparison between the XPSP and SCR methods reveals minor differences in zero-point offsets, typically within the range of 1 to 6\,mmag, indicating the accuracy of the re-calibration, and a two- to three-fold improvement in the zero-point precision. During this process, we also verified and corrected for the systematic errors related to CCD position. The corrected S-PLUS DR4 photometric data will provide a solid data foundation for conducting scientific research that relies on high-calibration precision.
Our results underscore the power of the XPSP method in combination with the SCR method, showcasing their effectiveness in enhancing calibration precision for wide-field surveys when combined with Gaia photometry and XP spectra, to be applied for other S-PLUS sub-surveys. 
\end{abstract}


\section{Introduction} \label{sec:intro}
Accurate and uniform photometric calibration presents a challenging task, yet is crucial for wide-field surveys due to rapid fluctuations in Earth's atmospheric opacity on time scales of seconds to minutes, instrumental effects (e.g., flat-field corrections), and electronics instability (e.g., variation in detector gain over time). Traditional optical photometric calibration relies on networks of standard stars with well-determined photometry, such as \cite{1992AJ....104..372L,2009AJ....137.4186L,2013AJ....146..131L} and \cite{2000PASP..112..925S}. However, the limited number of standard stars hinder traditional methods from meeting the calibration accuracy expectations of modern wide-field photometric surveys. Over the past two decades, significant advancements have been made in achieving high-precision calibration using various methods, broadly categorized into ``hardware-driven" and ``software-driven" approaches, as discussed by \cite{2022ApJS..259...26H}. Hardware-driven methods include the Ubercalibration method (\citealt{2008ApJ...674.1217P}), the Hypercalibration method (\citealt{2016ApJ...822...66F}), and the Forward Global Calibration Method \citep{2018AJ....155...41B}.  The software-driven approaches involve techniques such as the  Stellar Locus Regression method \citep{2009AJ....138..110H}, the Stellar Color Regression method (SCR; \citealt{2015ApJ...799..133Y}), and the Stellar Locus method \citep{2019A&A...631A.119L}.

The central idea of the SCR method is to predict the intrinsic colors of stars by utilizing stellar-atmospheric parameters, which has proven to be particularly effective in photometric 
re-calibration of wide-field surveys. For instance, when applied to the Sloan Digital Sky Survey (SDSS; \citealt{2000AJ....120.1579Y}) Stripe 82 (\citealt{2007AJ....134..973I}), it achieved a precision of 2--5\,mmag in the SDSS colors. Additionally, it has been employed for data from Gaia Data Release 2 and Early Data Release 3 (EDR3) to correct for magnitude/color-dependent systematic errors in the Gaia photometry (\citealt{2021ApJ...909...48N,2021ApJ...908L..14N}), yielding an unprecedented precision of 1\,mmag. 

\citet{2021ApJ...907...68H} utilized the SCR approach to re-calibrate the second data release (DR2) of the SkyMapper Southern Survey (SMSS; \citealt{2018PASA...35...10W}), revealing large zero-point offsets in the $uv$-bands. \cite{2022ApJS..259...26H} applied the method to SDSS Stripe 82 standard-star catalogs (\citealt{2007AJ....134..973I,2021MNRAS.505.5941T}), achieving a precision of 5\,mmag in the SDSS $u$-band and 2\,mmag in the $griz$-bands \citep{2015ApJ...799..133Y}). In addition, \cite{2022AJ....163..185X} and \cite{2023arXiv230805774X} applied the SCR method to the Pan-STARRS1 (PS1; \citealt{2012ApJ...750...99T}) data, effectively correcting for significant large-scale and small-scale spatial variations in the magnitude offsets and magnitude-dependent systematic errors. Other applications include Xiao et al. (2023, in prep), who use the SCR method to perform re-calibration on the J-PLUS DR3 photometric data, accurately measuring and correcting for the PS1 systematic errors and the metallicity-dependent systematic errors present in the J-PLUS DR3 photometric data. \cite{xiao} also performed the photometric calibration of Nanshan one-meter wide-field telescope $gri$-band imaging Data of the Stellar Abundance and Galactic Evolution Survey (SAGES; \citealt{Zheng18,Zheng19}) using the SCR method, achieving 1--2\,mmag precision in the zero-points.

Recently, the Gaia DR3 (\citealt{2021A&A...652A..86C,2022arXiv220800211G}) was released, which provides very low-resolution ($\lambda/\Delta \lambda\sim$ 50) XP spectra for roughly 220 million sources, with the majority having magnitudes $G < 17.65$. The XP spectra cover wavelengths from 336 to 1020\,nm, and have undergone precise internal \citep{2021A&A...652A..86C,2022arXiv220606143D} and external calibrations \citep{2022arXiv220606205M}. Unfortunately, the Gaia XP spectra exhibit systematic errors that depend on magnitude and color, particularly at wavelengths below 400\,nm (see, e.g., \citealt{2022arXiv220606205M, huang}). 

More recently, comprehensive corrections to the Gaia XP spectra have been provided by \cite{huang}, utilizing spectra from CALSPEC \citep{CALSPEC14, CALSPEC22} and Hubble's Next Generation Spectral Library (NGSL; \citealt{NGSL}). In this process, the spectroscopy-based SCR method \citep{2015ApJ...799..133Y} was employed as well. Based on the corrected Gaia XP spectra, Xiao et al. (2023, in prep) further develop the XP spectra-based photometric synthesis (XPSP, hereafter) method, and applied it to the photometric calibration of J-PLUS DR3 data. The consistency between the J-PLUS zero-points predicted by the XPSP method after XP spectra correction and the SCR method is better than 5\,mmag, which represents a twofold improvement compared to the consistency between the 
J-PLUS zero-points predicted by the XPSP method with uncorrected XP spectra and the SCR method.

Located at the Cerro Tololo Interamerican Observatory, the Southern Photometric Local Universe Survey (S-PLUS\footnote{\url{http://splus.iag.usp.br}}; \citealt{2019MNRAS.489..241M}) employs a 83\,cm telescope to obtain images on a single CCD. The photometric calibration of S-PLUS DR4 is carried out using photometric data from GALEX, SDSS, Pan-Starrs, Skymapper, and so on, along with the spectral energy distribution (SED) information for calibration sources \citep{2022MNRAS.511.4590A}. 
 However, this method i) relies on reference catalogs that do not have uniform calibration precision across the S-PLUS footprint; ii) this approach relies on synthetic stellar models, and will inherit any systematic errors present in these (for instance, \citealt{2022MNRAS.511.4590A} observe zero-point offsets as high as 50\,mmag for J0395 just by changing the synthetic spectral library); and iii) it relies on \cite{1998ApJ...500..525S} extinction maps, and thus fails at low Galactic latitudes and exhibits spatially-dependent systematic errors, up to 0.02\,mag \citep{2022ApJS..260...17S}; iv) and aperture corrections for the determination of aperture magnitudes.
Improvement of the photometric calibration of S-PLUS is crucial, given the importance of high-precision investigations, in particular those that seek accurate determinations of stellar parameters and elemental abundances. 

In this study, we utilize both an improved XPSP method and the SCR method to conduct photometric re-calibration of the S-PLUS DR4 data (Herpich et al., in prep.), aiming to achieve uniform photometry with accuracy better than 1\%. The structure of this paper is as follows. We present the data used in this work in Section\,\ref{sec:data}. 
The predictions of S-PLUS magnitudes with the XPSP method and the SCR method are presented in Section\,\ref{sec:method}, followed by a description of the systematic errors presented in S-PLUS DR4 data in Section\,\ref{sec:res}.
A discussion is carried out in Section\,\ref{sec:discussion}. Finally, we provide brief conclusions in Section\,\ref{sec:conclusion}.

\begin{figure*}[ht!] \centering
\resizebox{\hsize}{!}{\includegraphics{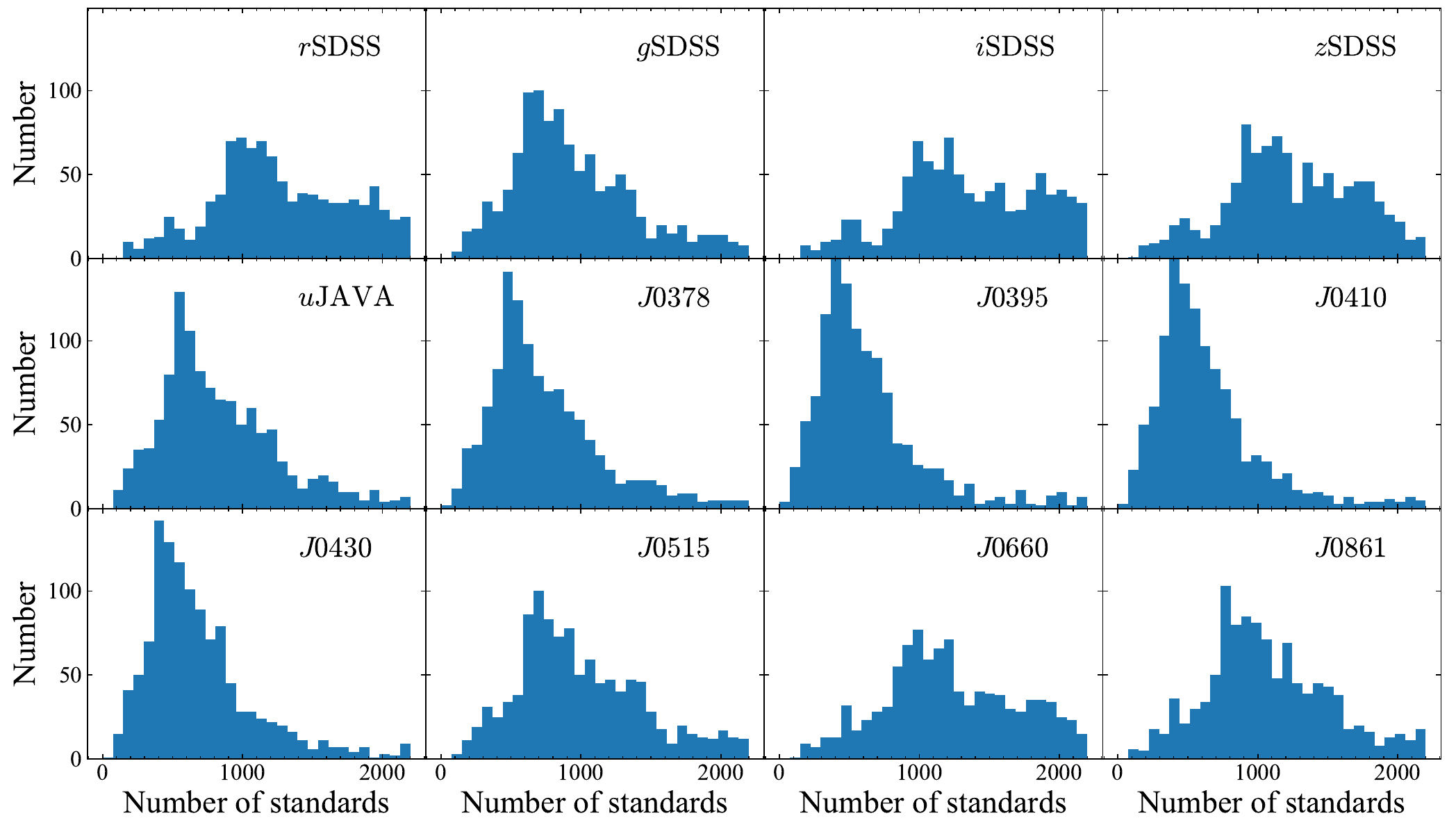}}
\caption{{\small Histograms of the number of standard stars for the XPSP method in each image. The bands are labeled in each panel.
}}
\label{Fig:standars}
\end{figure*}

\section{Data} \label{sec:data}
\subsection{S-PLUS Data Release 4} \label{sec:ps1}
The S-PLUS DR4 encompasses 1629 pointings, covering approximately 3000 deg$^2$ of the Southern sky, including the Main Survey with PStotal and PSF photometry, the Magellanic Clouds (MCs) with PStotal and PSF photometry, and the Disk Survey with PSF photometry (Herpich et al., in prep). The PStotal photometry was the one used for the calibration, and is the best representation for the total magnitude of a point source in the S-PLUS catalogs (for the aperture photometry). The S-PLUS data were obtained using the T80-South telescope\footnote{\url{https://noirlab.edu/public/programs/ctio/t80-south-telescope/}}. The panoramic camera features a single charge-coupled device (CCD) with a resolution of 9.2k $\times$ 9.2k pixels, a field of view (FoV) measuring $1.4^{\circ}\times 1.4^{\circ}$, and a pixel scale of 0.55$^{\prime\prime}$pix$^{-1}$ \citep{2015IAUGA..2257381M}. It employs 5 broad-band filters ($u{\rm JAVA}$, $g{\rm SDSS}$, $r{\rm SDSS}$, $i{\rm SDSS}$, and $z{\rm SDSS}$) and 7 medium-band filters ($J0378$, $J0395$, $J0410$, $J0430$, $J0515$, $J0660$, and $J0861$) within the optical range. It is essential to note that the S-PLUS DR4 magnitudes mentioned in this paper refer to the magnitudes calibrated following \cite{2022MNRAS.511.4590A}.

\subsection{Gaia Early Data Release 3} \label{sec:gaia}
The Gaia EDR3 (\citealt{2021A&A...649A...1G,2021A&A...650C...3G}) provides the most precise photometric data available to date for approximately 1.8 billion stars. The magnitudes in the $G$, $G_{\rm BP}$, and $G_{\rm RP}$ bands have been uniformly calibrated with accuracy at the mmag level \citep[e.g., ][]{2021ApJS..255...20A,2021ApJ...922..211N}. To address magnitude-dependent systematic errors, which are estimated to be around 1\% in these bands for Gaia EDR3, \citet{2021ApJ...908L..24Y} utilized approximately 10,000 Landolt standard stars from \cite{2013AJ....146...88C}. In our study, we adopt the magnitudes of $G$, $G_{\rm BP}$, and $G_{\rm RP}$ as corrected by \citet{2021ApJ...908L..24Y} by default.

\subsection{Gaia Data Release 3} \label{sec:gaia2}
Gaia DR3 (\citealt{2021A&A...652A..86C,2022arXiv220800211G}), based on 34 months of observations, provides very low-resolution ($\lambda/\Delta \lambda\sim$ 50) XP spectra for approximately 220 million sources, with the majority having magnitudes $G < 17.65$. The XP spectra cover a wavelength range from 336 to 1020\,nm, and have undergone precise internal calibrations \citep{2021A&A...652A..86C,2022arXiv220606143D} as well as external calibrations \citep{2022arXiv220606205M}. However, it is crucial to note that Gaia XP spectra are subject to systematic errors that depend on magnitude, color, and extinction, especially at wavelengths below 400\,nm \citep[see][]{2022arXiv220606205M, huang}. A comprehensive set of corrections, based on reference spectra from CALSPEC and NGSL, have been provided by \cite{huang}. In this paper, the term ``corrected Gaia XP spectra" refers to the Gaia XP spectra as rectified by \cite{huang}.

\subsection{LAMOST Data Release 7} \label{sec:lm}
The Large Sky Area Multi-Object Fiber Spectroscopic Telescope (LAMOST; \citealt{2012RAA....12.1197C}; \citealt{2012RAA....12..735D}; \citealt{2012RAA....12..723Z}; \citealt{2014IAUS..298..310L}) is a quasi-meridian reflecting Schmidt telescope equipped with 4000 fibers and a field-of-view spanning 20\,deg$^2$. LAMOST's Data Release 7 (DR7; \citealt{2015RAA....15.1095L}) presents a comprehensive data set comprising 10,640,255 low-resolution spectra, over the full optical wavelength range from 369 to 910\,nm, with a spectral resolution of $R \approx 1800$. To derive fundamental stellar parameters, including effective temperature ($T_{\rm eff}$), surface gravity ($\log g$), and metallicity ($\rm [Fe/H]$), the LAMOST Stellar Parameter Pipeline (LASP; \citealt{2011RAA....11..924W}) has been employed. The internal precision typically attained for these parameters is approximately 110\,K for $T_{\rm eff}$, 0.2\,dex for $\log g$, and 0.1\,dex for ${\rm [Fe/H]} \gtrsim -2.5$ \citep{2015RAA....15.1095L}.

\section{Predictions of S-PLUS Magnitudes}
\label{sec:method}

\begin{figure*}[ht!] \centering
\resizebox{\hsize}{!}{\includegraphics{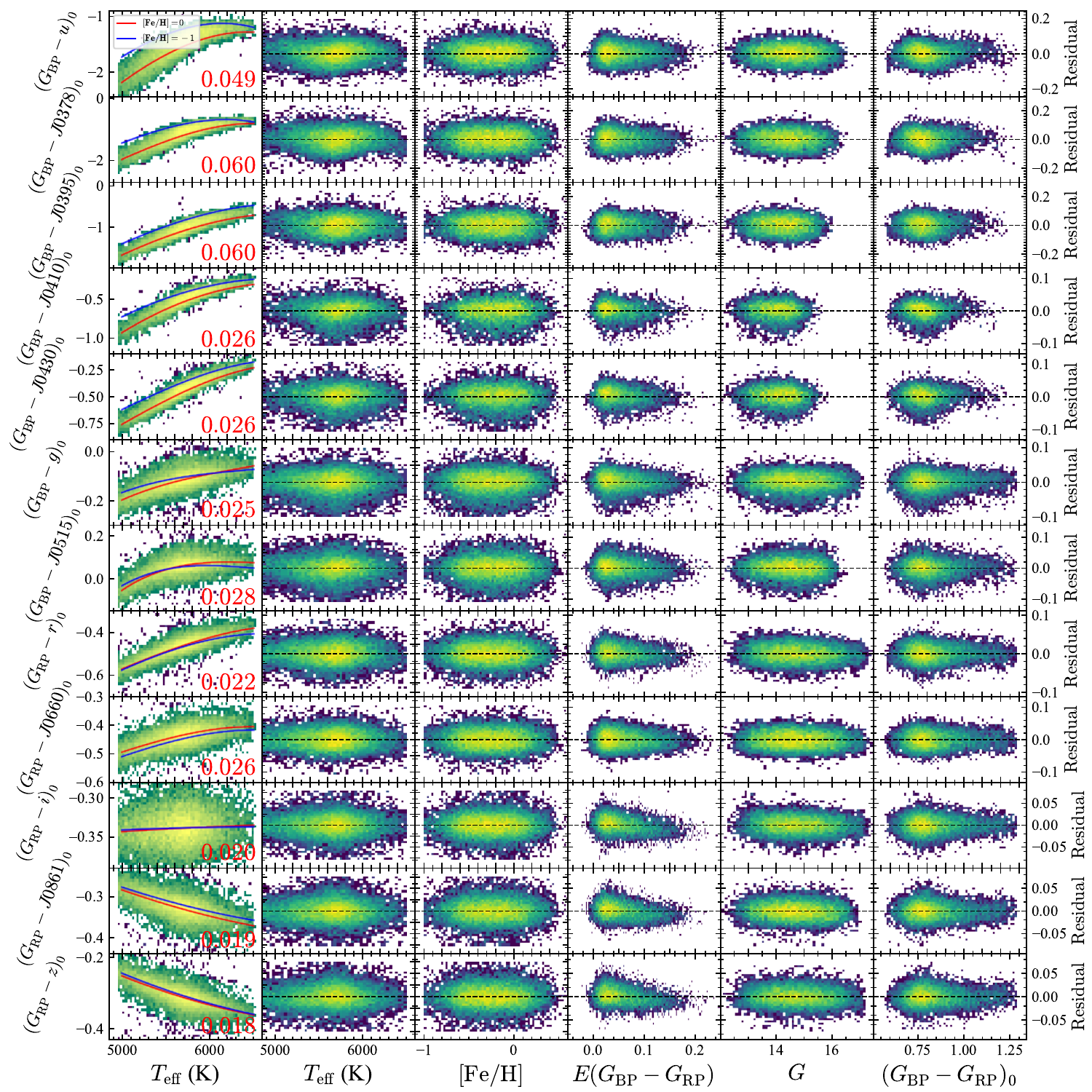}}
\caption{{\small Two-dimensional polynomial fitting of intrinsic colors with respect to $T_{\rm eff}$ and $\rm [Fe/H]$ for the calibration stars in the SCR method. The intrinsic colors include $G_{\rm BP}-u{\rm JAVA}$, $G_{\rm BP}-J0378$, $G_{\rm BP}-J0395$, $G_{\rm BP}-J0410$, $G_{\rm BP}-J0430$, $G_{\rm BP}-g{\rm SDSS}$, $G_{\rm BP}-J0515$, $G_{\rm RP}-r{\rm SDSS}$, $G_{\rm RP}-J0660$, $G_{\rm RP}-i{\rm SDSS}$, $G_{\rm RP}-J0861$, and $G_{\rm RP}-z{\rm SDSS}$.  From left to right, the fit results after 3$\sigma$ clipping are shown in the first column, with the red and blue curves representing results for $\rm [Fe/H]=0$ and [Fe/H] = $=-1$, respectively. The fitting residuals are labeled in red. In the second to sixth columns, the residuals are plotted against $T_{\rm eff}$, $\rm [Fe/H]$, extinction of $E(G_{\rm BP}-G_{\rm RP})$, $G$ magnitude, and $(G_{\rm BP}-G_{\rm RP})_0$ color, respectively. Zero residuals are denoted by black dotted lines.
}}
\label{Fig:scr}
\end{figure*}

\begin{deluxetable*}{ccccccccccc}[ht!]
\tablecaption{Coefficients used to Obtain Intrinsic Colors as Functions of $T_{\rm eff}$ and ${\rm [Fe/H]}$ in the 12 bands. In the table, the symbol $ei$ represents $10^{-i}$.
$C^{\rm mod}_{\rm 0}={a_0}\cdot x^3+{a_1}\cdot y^3+{a_2}\cdot x^2\cdot y+{a_3}\cdot x\cdot y^2+{a_4}\cdot x^2+{a_5}\cdot y^2+{a_6}\cdot x\cdot y+{a_7}\cdot x+{a_8}\cdot y+{a_9}$, where, $x$ is $T_{\rm eff}$ and $y$ is ${\rm [Fe/H]}$. \\
 \label{tab:1}}
\tablehead{
\colhead{Intrinsic Color} & \colhead{$a_0$} & \colhead{$a_1$} & \colhead{$a_2$} & \colhead{$a_3$} & \colhead{$a_4$} & \colhead{$a_5$} & \colhead{$a_6$} & \colhead{$a_7$} & \colhead{$a_8$} & \colhead{$a_9$}}
\startdata
$(G_{\rm BP}-u{\rm JAVA})_{\rm 0}$ & $-$6.140e11 & $+$0.006 & $+$5.692e8 & $+$1.269e5 & $+$6.605e7 & $-$1.901e1 & $-$3.964e4 & $-$8.742e4 & $-$0.028 & $-$6.650 \\
$(G_{\rm RP}-J0378)_{\rm 0}$ & $-$1.392e10 & $+$0.023 & $+$9.537e8 & $-$1.723e5 & $+$1.984e6 & $+$7.471e5 & $-$8.069e4 & $-$8.177e3 & $+$1.026 & $+$6.731 \\
$(G_{\rm RP}-J0395)_{\rm 0}$ & $-$2.697e11 & $+$0.075 & $-$1.265e8 & $-$1.121e5 & $+$2.235e7 & $+$1.053e1 & $+$0.0002 & $+$7.561e4 & $-$0.746 & $-$7.669 \\
$(G_{\rm RP}-J0410)_{\rm 0}$ & $-$1.376e12 & $-$0.002 & $-$3.555e8 & $+$6.523e5 & $-$1.614e7 & $-$4.446e1 & $+$5.634e4 & $+$2.398e3 & $-$2.244 & $-$8.709 \\
$(G_{\rm RP}-J0430)_{\rm 0}$ & $-$3.228e11 & $+$0.022 & $-$9.292e9 & $-$8.655e6 & $+$4.584e7 & $+$5.971e2 & $+$1.536e4 & $-$1.705e3 & $-$0.676 & $+$0.345 \\
$(G_{\rm BP}-g{\rm SDSS})_{\rm 0}$ & $+$1.745e11 & $+$0.004 & $+$2.991e9 & $-$4.360e6 & $-$3.311e7 & $+$1.609e2 & $-$9.492e6 & $+$2.158e3 & $-$0.067 & $-$4.891 \\
$(G_{\rm RP}-J0515)_{\rm 0}$ & $+$3.762e11 & $+$0.007 & $-$8.821e9 & $+$2.966e6 & $-$7.439e7 & $-$2.790e2 & $+$1.346e4 & $+$4.886e3 & $-$0.493 & $-$10.589 \\
$(G_{\rm RP}-r{\rm SDSS})_{\rm 0}$ & $-$9.563e12 & $+$0.018 & $+$1.687e8 & $-$1.396e5 & $+$1.278e7 & $+$9.315e2 & $-$0.0002 & $-$3.890e4 & $+$0.555 & $-$0.624 \\
$(G_{\rm RP}-J0660)_{\rm 0}$ & $-$8.887e12 & $+$0.015 & $+$1.730e9 & $-$1.054e5 & $+$1.206e7 & $+$7.299e2 & $-$3.287e5 & $-$4.427e4 & $+$0.142 & $-$0.185 \\
$(G_{\rm RP}-i{\rm SDSS})_{\rm 0}$ &  &  &  &  & $-$1.038e9 & $-$3.942e3 & $+$2.666e6 & $+$1.701e5 & $-$0.019 & $-$0.402 \\
$(G_{\rm BP}-J0861)_{\rm 0}$ &  &  &  &  & $+$1.110e8 & $-$3.099e4 & $-$4.378e6 & $-$1.853e4 & $+$0.015 & $+$0.365 \\
$(G_{\rm RP}-z{\rm SDSS})_{\rm 0}$ &  &  &  &  & $+$1.630e8 & $-$1.335e3 & $+$3.958e6 & $-$.853e4 & $-$0.029 & $+$0.639 \\
\enddata
\end{deluxetable*}

In this section, we describe how to obtain the predicted magnitudes for the 12 photometric bands of S-PLUS using the improved XPSP method and SCR method.

\subsection{XPSP Method with Corrected Gaia XP Spectra} \label{xpsp}

The synthetic photometry method involves projecting the Spectral Energy Distribution (SED) at the top of atmosphere of a source onto the transmission curve of the photometric system. Following by Xiao et al. (in prep.), we compute the synthetic magnitude in the AB system \citep{1983ApJ...266..713O,1996AJ....111.1748F} for each S-PLUS band.

To account for the $u{\rm JAVA}$-band's wavelength range (322 to 382\,nm), slightly bluer than that of the Gaia XP spectra (336 to 1020\,nm), we perform numerical extrapolation to extend the Gaia XP spectra. For each source, we obtain a linear function for the Gaia XP spectra flux density with wavelength through fitting of the Gaia XP spectral data over the range of 336\,nm to 382\,nm for individual stars. 
This approach has been proposed and validated in the process of re-calibration of J-PLUS photometry, after evaluating multiple extrapolation methods (Xiao et al. in prep.).

In this study, we select calibration samples satisfying the following constraints: magnitude errors are less than 0.02\,mag for the $u{\rm JAVA}$, $J0378$, and $J0395$ bands, the bluest ones, and less than 0.01\,mag for the others. Consequently, we obtain 1,522,862, 1,319,587, 1,002,181, 597,486, 696,915, 2,567,317, 1,325,138, 3,843,548, 3,636,016, 3,984,143, 2,687,153, and 3,391,692 calibration stars in the $u{\rm JAVA}$, $J0378$, $J0395$, $J0410$, $J0430$, $g{\rm SDSS}$, $J0515$, $r{\rm SDSS}$, $J0660$, $i{\rm SDSS}$, $J0861$, and $z{\rm SDSS}$ bands, respectively.  We conduct a count of the standard stars on each image, and present histograms of their distribution in Figure\,\ref{Fig:standars}.

\begin{figure*}[ht!] \centering
\resizebox{\hsize}{!}{\includegraphics{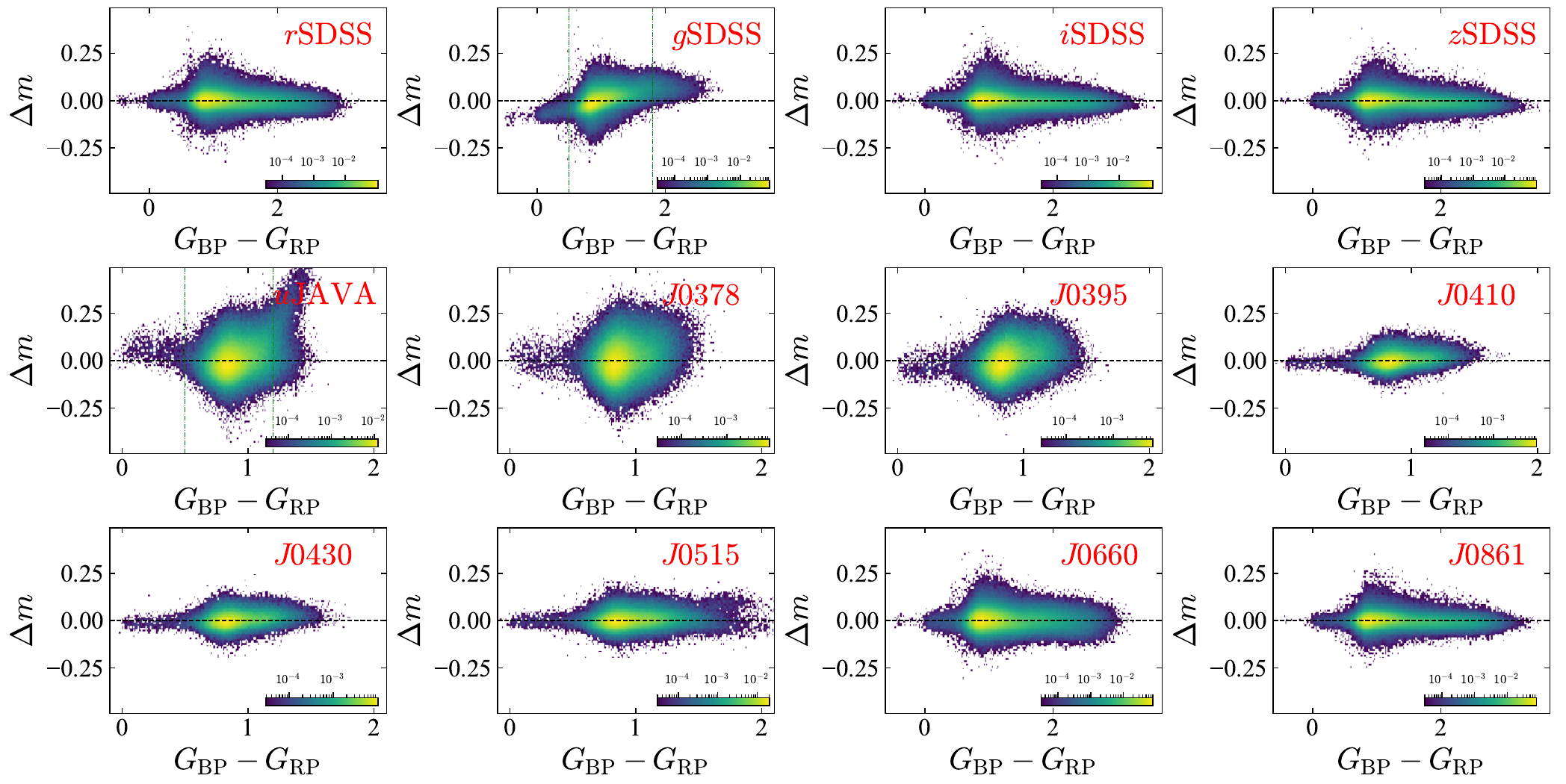}}
\caption{{\small Magnitude offsets between the XPSP predicted magnitudes and the S-PLUS magnitudes, as a function of $G_{\rm BP}-G_{\rm RP}$ color, for all 12 bands. The 
colors represents the density of points, and the bands are indicated in each panel. Zero residuals and two color cuts are denoted by the black and green dotted lines, respectively. Color bars are plotted in the lower right corner for each panel.
}}
\label{Fig:resi_bprp}
\end{figure*}

\subsection{SCR Method with Gaia Photometry and LAMOST Spectra}
To obtain a cross-validation of the calibration results obtained by the XPSP method, we also employ the independent SCR method to re-calibrate the Main Survey sample.

The SCR method comprises two key techniques: intrinsic color prediction and reddening correction. The former can be performed based on either spectroscopic or photometric data, while the latter necessitates precise measurement of the reddening coefficients relative to extinction values. The SCR method typically involves defining the relationship between the intrinsic colors and the physical quantities using a sample of low-extinction stars, which is then applied to the entire sample to obtain predicted magnitudes. A detailed description of the SCR method is as follows.

First, we select calibration samples with the same constraints in photometric re-calibration of J-PLUS as Xiao et al. (in prep.). Ultimately, we obtain 13,869, 12,714, 10,954, 8,814, 9,873, 16,301, 14,049, 16,558, 16,522, 16,464, 15,974, and 16,402 calibration stars in the $u{\rm JAVA}$, $J0378$, $J0395$, $J0410$, $J0430$, $g{\rm SDSS}$, $J0515$, $r{\rm SDSS}$, $J0660$, $i{\rm SDSS}$, $J0861$, and $z{\rm SDSS}$ bands, respectively. Then, we consider twelve colors for the twelve S-PLUS bands when combined with Gaia photometry, denoted as ${\boldsymbol C} = {\boldsymbol G}_{\rm BP/RP}-{\boldsymbol m}_{\rm SPLUS}$ $=$ ($G_{\rm BP}-u{\rm JAVA}$, $G_{\rm BP}-J0378$, $G_{\rm BP}-J0395$, $G_{\rm BP}-J0410$, $G_{\rm BP}-J0430$, $G_{\rm BP}-g{\rm SDSS}$, $G_{\rm BP}-J0515$, $G_{\rm RP}-r{\rm SDSS}$, $G_{\rm RP}-J0660$, $G_{\rm RP}-i{\rm SDSS}$, $G_{\rm RP}-J0861$, $G_{\rm RP}-z{\rm SDSS}$). 

To correct for reddening, we adopted the same process as \cite{2022AJ....163..185X} and Xiao et al. (in prep.). We adopt the values of $E(G_{\rm BP}-G_{\rm RP})$ obtained with the star-pair method \citep{2013MNRAS.430.2188Y,2020ApJ...905L..20R}. And, the reddening coefficients with respect to $E(G_{\rm BP}-G_{\rm RP})$ for the 12 colors constructed by Yuan et al. (in prep.).

For each color, we fit the intrinsic color as a function of $T_{\rm eff}$ and $\rm [Fe/H]$ using a two-dimensional polynomial. Specifically, we use second-order polynomials for the $G_{\rm RP}-i$, $G_{\rm RP}-J0861$, and $G_{\rm RP}-z$ colors, and third-order polynomials for the other colors. The intrinsic colors ($\boldsymbol {C_{\rm 0}}$) can then be estimated using observed colors ${\boldsymbol C}$ minus the product of reddening coefficients and extinction $E(G_{\rm BP}-G_{\rm RP})$.

The fitting results of the intrinsic colors as a function of $T_{\rm eff}$, $\rm [Fe/H]$, and extinction of $E(G_{\rm BP}-G_{\rm RP})$ are shown in Figure\,\ref{Fig:scr}, and the corresponding fitting parameters are listed in Table\,\ref{tab:1}. The intrinsic-color fitting residuals are, respectively, 49, 60, 60, 26, 26, 25, 28, 22, 26, 20, 19, and 18\,mmag for $G_{\rm BP}-u{\rm JAVA}$, $G_{\rm BP}-J0378$, $G_{\rm BP}-J0395$, $G_{\rm BP}-J0410$, $G_{\rm BP}-J0430$, $G_{\rm BP}-g{\rm SDSS}$, $G_{\rm BP}-J0515$, $G_{\rm RP}-r{\rm SDSS}$, $G_{\rm RP}-J0660$, $G_{\rm RP}-i{\rm SDSS}$, and $G_{\rm RP}-J0861$ colors, suggesting that S-PLUS magnitudes can be predicted for individual stars with a precision of 20 to 60\,mmag using the Gaia and LAMOST data. Furthermore, the fitting residuals exhibit no dependence on $T_{\rm eff}$, $\rm [Fe/H]$, and $E(G_{\rm BP}-G_{\rm RP})$.

Having obtained the intrinsic-color fitting functions, we apply them to the calibration stars to obtain the derived magnitudes $m_{\rm SCR}$ for each image using Equation\,\ref{e2}:
    \begin{eqnarray}
    \begin{aligned}
    {\boldsymbol m_{\rm SCR}}=~&{\boldsymbol G}_{\rm BP,RP}-{\boldsymbol C_{\rm 0}^{\rm mod}}(T_{\rm eff},~\rm [Fe/H])-\\
    &{\boldsymbol R} \times E(G_{\rm BP}-G_{\rm RP})~. \label{e2}
    \end{aligned}
    \end{eqnarray}

\begin{figure*}[ht!] \centering
\resizebox{\hsize}{!}{\includegraphics{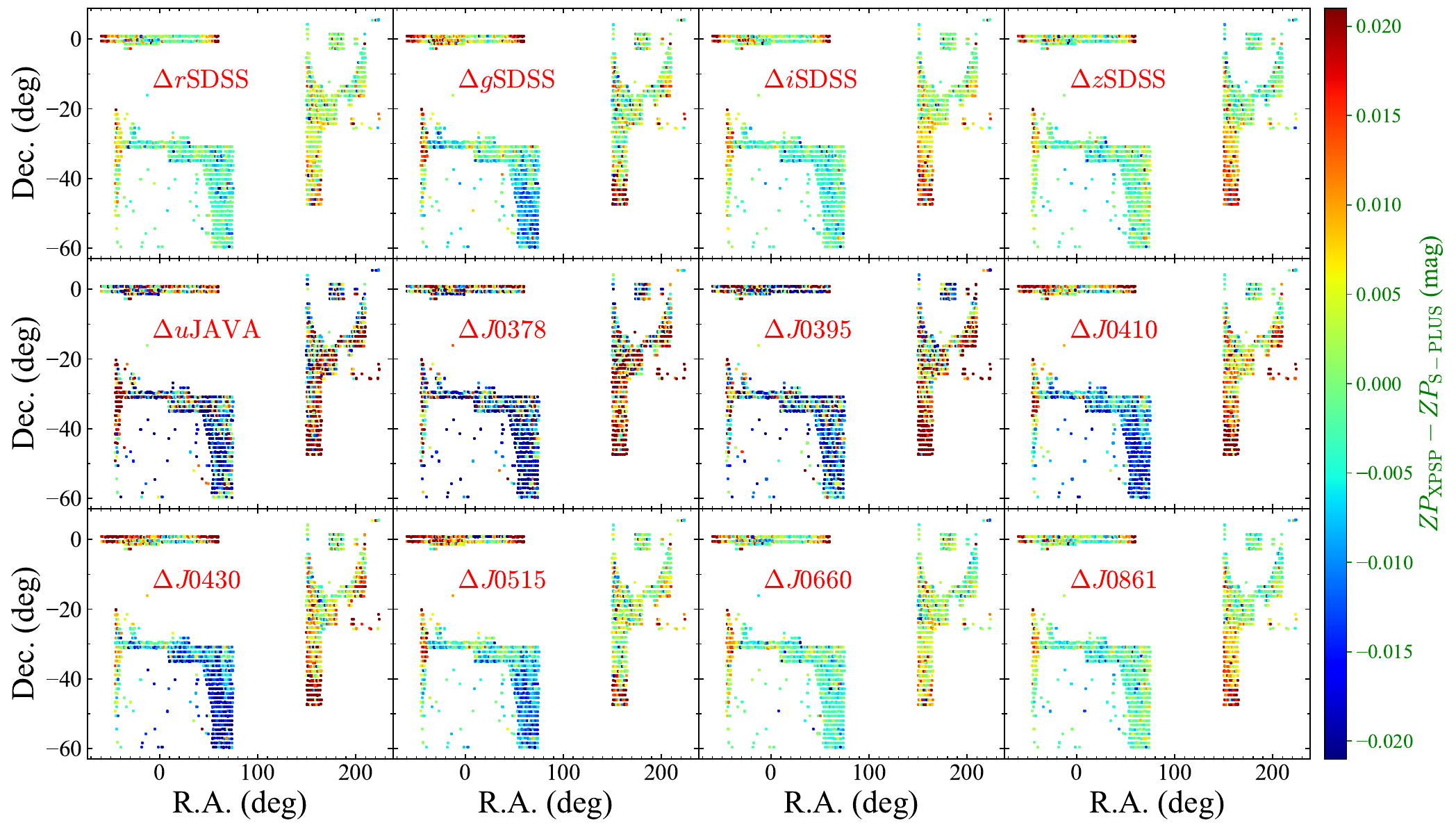}}
\caption{{\small Spatial variations of the difference between the XPSP zero-points are applied, and the S-PLUS zero-points in each image of the Main Survey, using PStotal photometry from S-PLUS are shown. The bands are marked in each panel in red, and the color bar indicating the differences is shown on the right. 
}}
\label{Fig:spatial_dis_pstotal_main}
\end{figure*}

\begin{figure*}[ht!] \centering
\resizebox{\hsize}{!}{\includegraphics{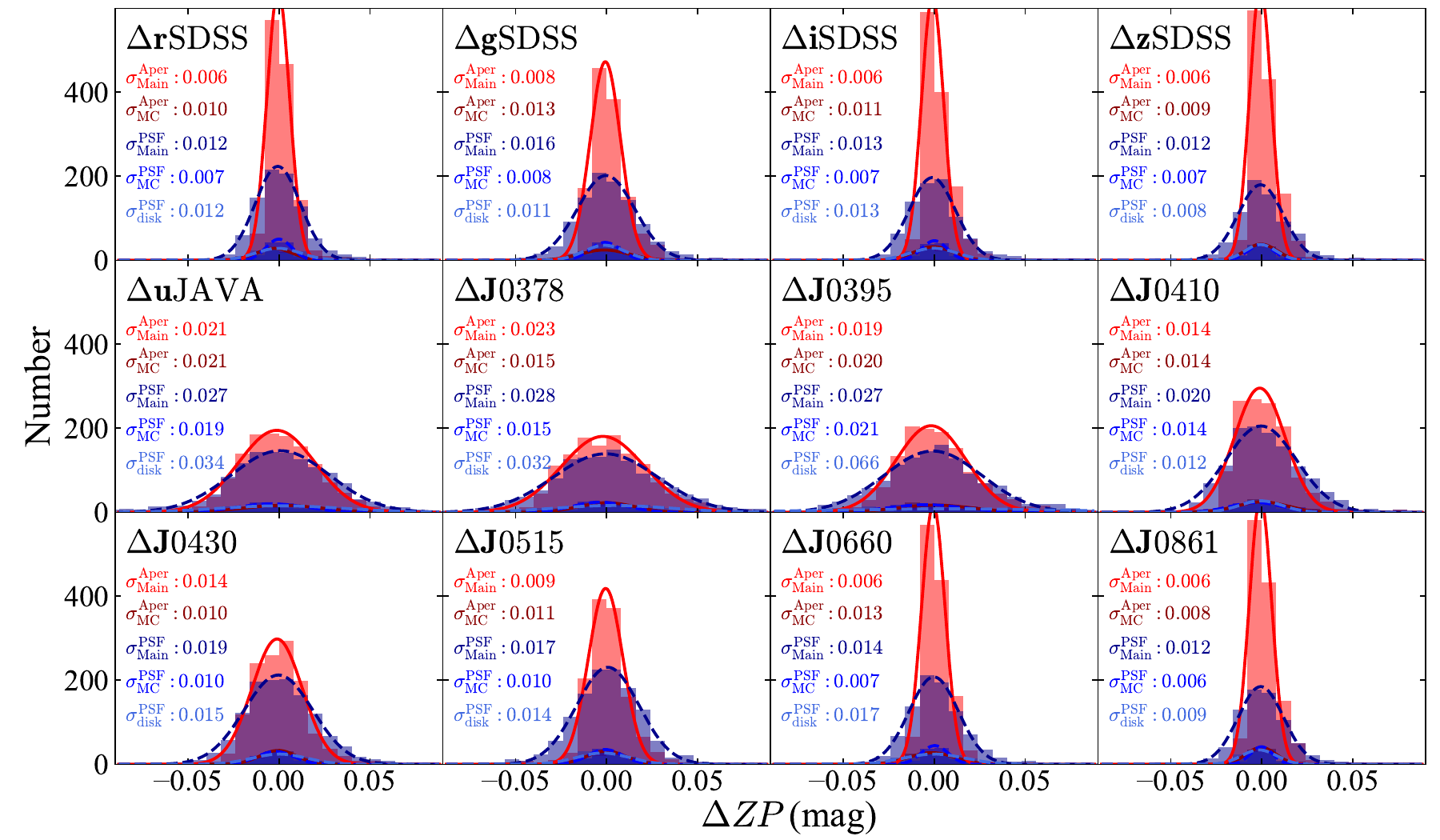}}
\caption{{\small Histograms of the residuals in zero-points between the XPSP magnitudes are applied, along with the S-PLUS magnitudes. The results shown are for the Main Survey with PStotal photometry (dark red), the MCs with PStotal photometry (red), the Main Survey with PSF photometry (dark blue), the MCs with PSF photometry (blue), and the Disk Region with PSF photometry (light blue).  The bands are labeled in the top-left corners of each panel.
Gaussian-fitting results are plotted with the same colors, using solid and dashed curves for PStotal photometry and PSF photometry, respectively. The sigma values are labeled in the top-left corners with the same colors.
}}
\label{Fig:hist}
\end{figure*}

\begin{figure*}[ht!] \centering
\resizebox{\hsize}{!}{\includegraphics{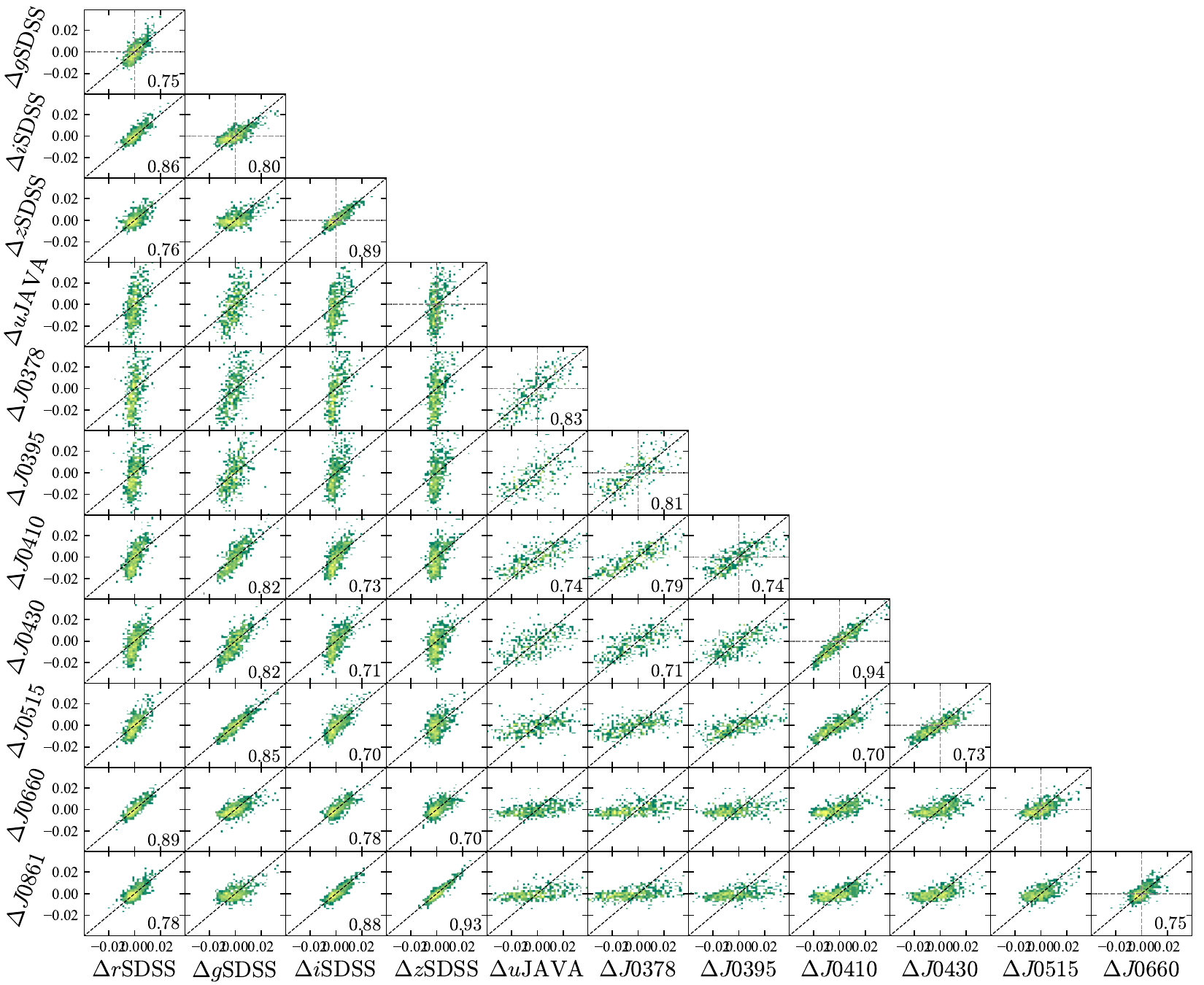}}
\caption{{\small Correlation plots between the zero-point offsets. The correlation coefficients are shown when they have values greater than 0.7. The colors in each panel indicates the number density of stars, and the dashed black line represents the one-to-one line.
}}
\label{Fig:r}
\end{figure*}

\begin{figure*}[ht!] \centering
\resizebox{\hsize}{!}{\includegraphics{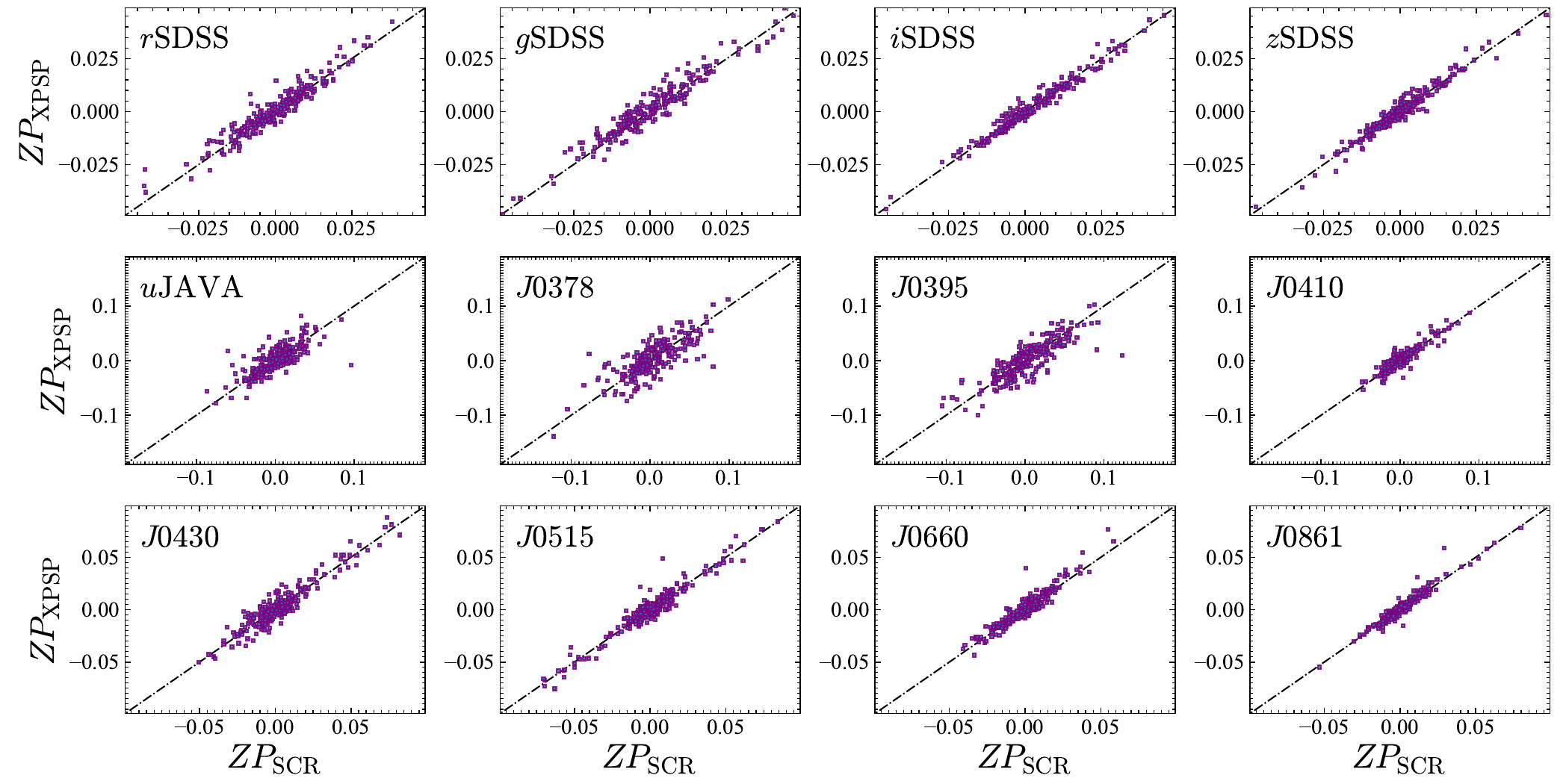}}
\caption{{\small Comparison of the zero-points for each of the two methods based on stars in common between the XPSP and the SCR methods samples for all 12 bands. The dashed black line represents the one-to-one line.}}
\label{Fig:1v1}
\end{figure*}

\begin{figure*}[ht!] \centering
\resizebox{\hsize}{!}{\includegraphics{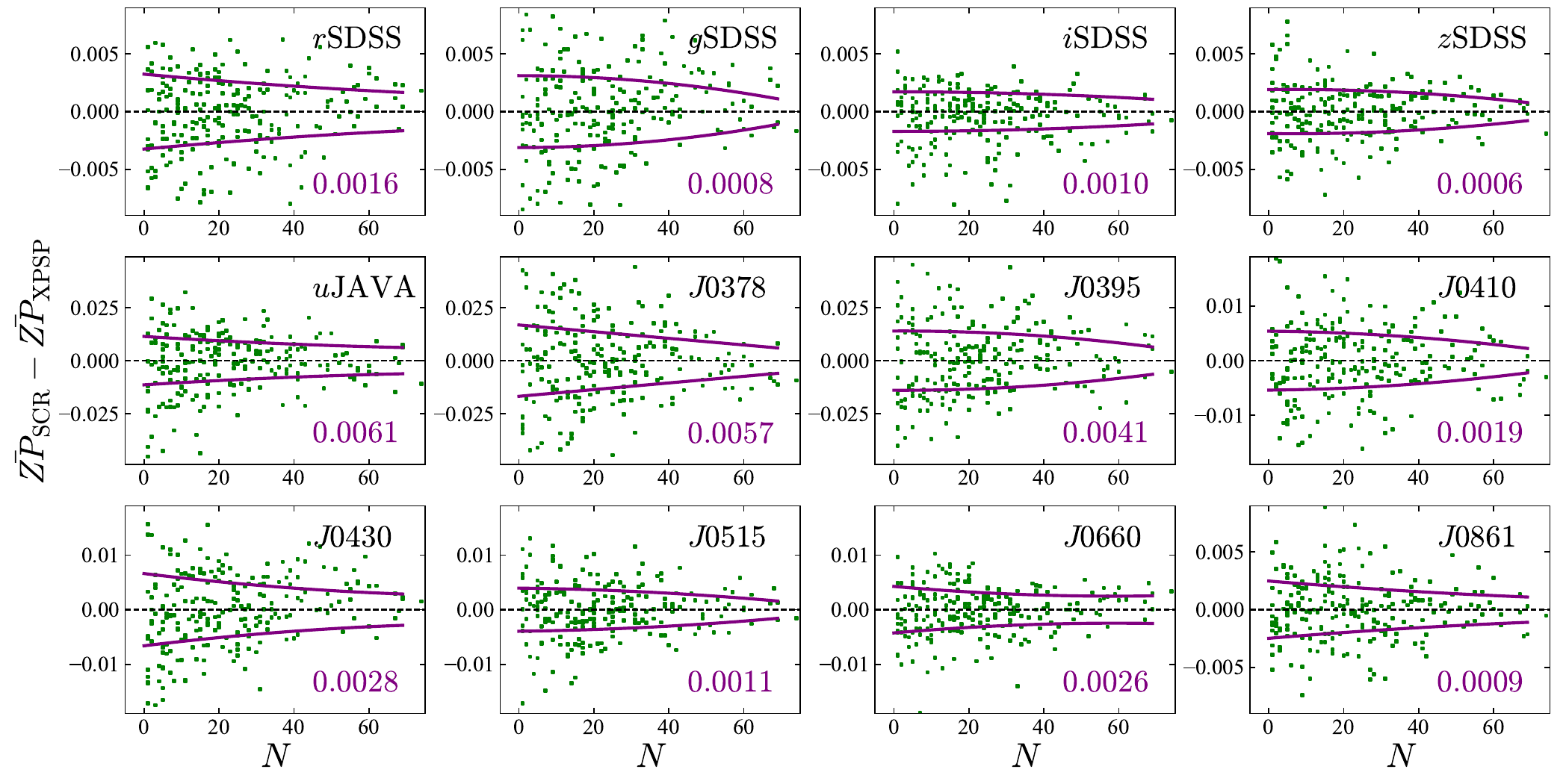}}
\caption{{\small Comparison of the zero-points for each of the two methods based on stars in common between the XPSP and the SCR methods samples for all 12 bands, as a function of star numbers in each 
image. The bands are marked in each panel. The green 
points show the difference of the zero-points, and their standard deviations are indicated by red curves. The standard deviations for 
$N = 60$ stars are labeled in each panel.
}}
\label{Fig:precision}
\end{figure*}

\begin{figure*}[ht!] \centering
\resizebox{\hsize}{!}{\includegraphics{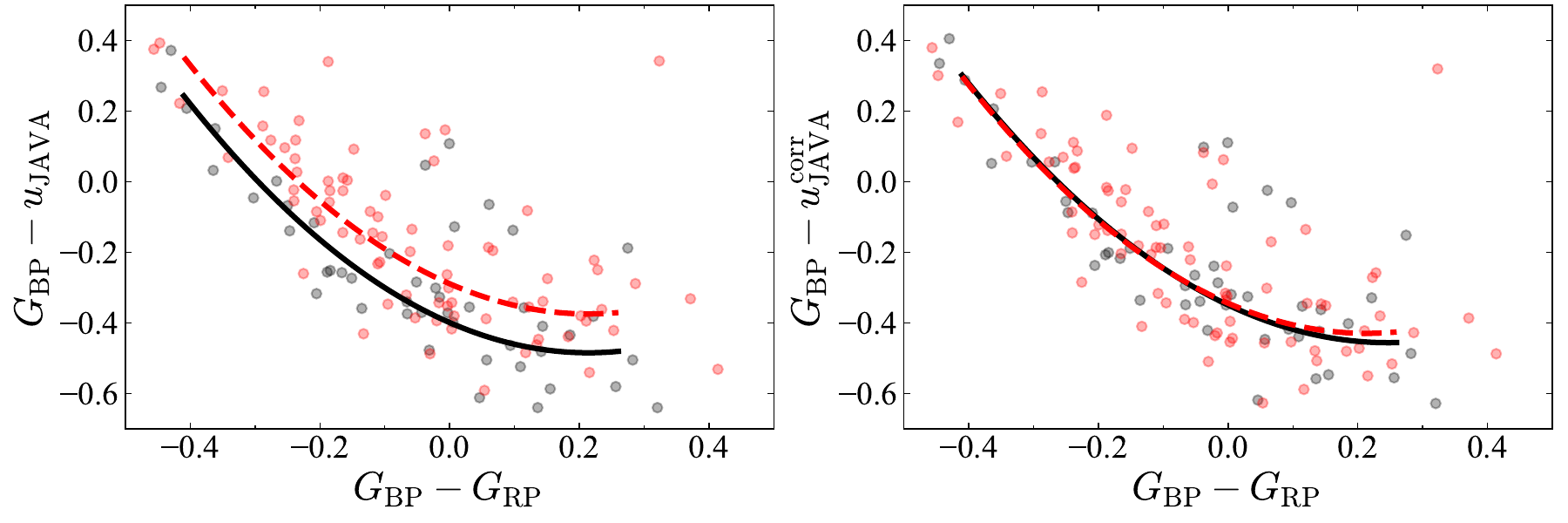}}
\caption{{\small The WD loci $G-u{\rm JAVA}$ vs. $G_{\rm BP}-G_{\rm RP}$ before (left panel) and after (right) re-calibration in the $u{\rm JAVA}$-band. The red and black points represent the $R.A.<90^{\circ}$ and $decl.<-40^{\circ}$ region and the $R.A.>180^{\circ}$ and $decl.>-30^{\circ}$ region, respectively. The red-dotted curve and black curve correspond to the quadratic polynomial fitting results for the red dots and black dots, respectively.
}}
\label{Fig:check_wd}
\end{figure*}

\begin{figure*}[ht!] \centering
\resizebox{\hsize}{!}{\includegraphics{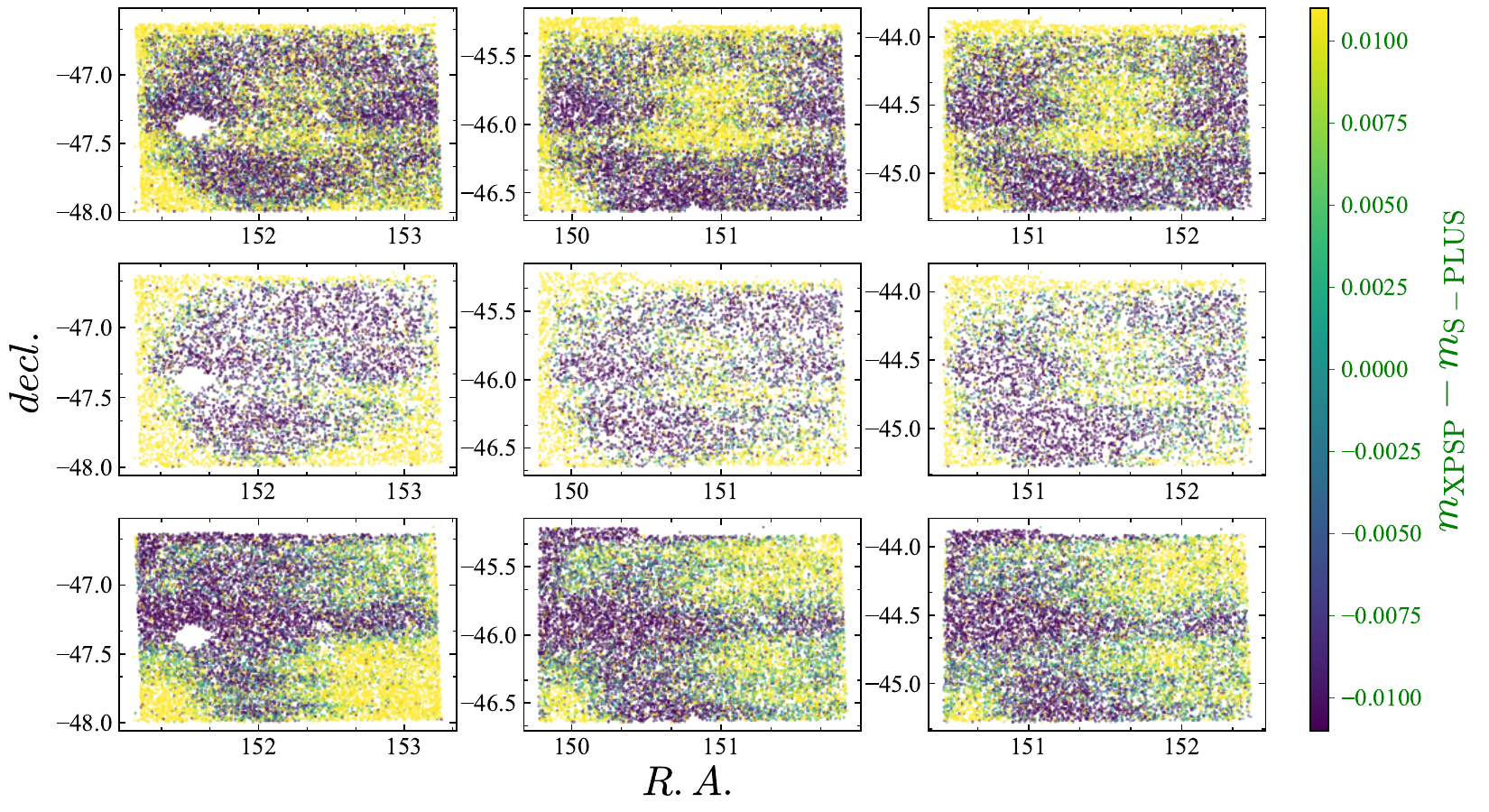}}
\caption{{\small An example showing the spatial distribution over the CCD of the difference between the XPSP method and the S-PLUS magnitudes. From left to right, the results are shown for the $g{\rm SDSS}$, $J0515$, and $J0861$ bands, respectively. From top to bottom, three different observations are shown with $\texttt{tile\_id}$ of $\texttt{HYDRA-0161}$, $\texttt{HYDRA-0152}$, and $\texttt{HYDRA-0145}$. A color bar is shown on the right.}
}
\label{Fig:mxpsp_mplus}
\end{figure*}

\begin{figure*}[ht!] \centering
\resizebox{\hsize}{!}{\includegraphics{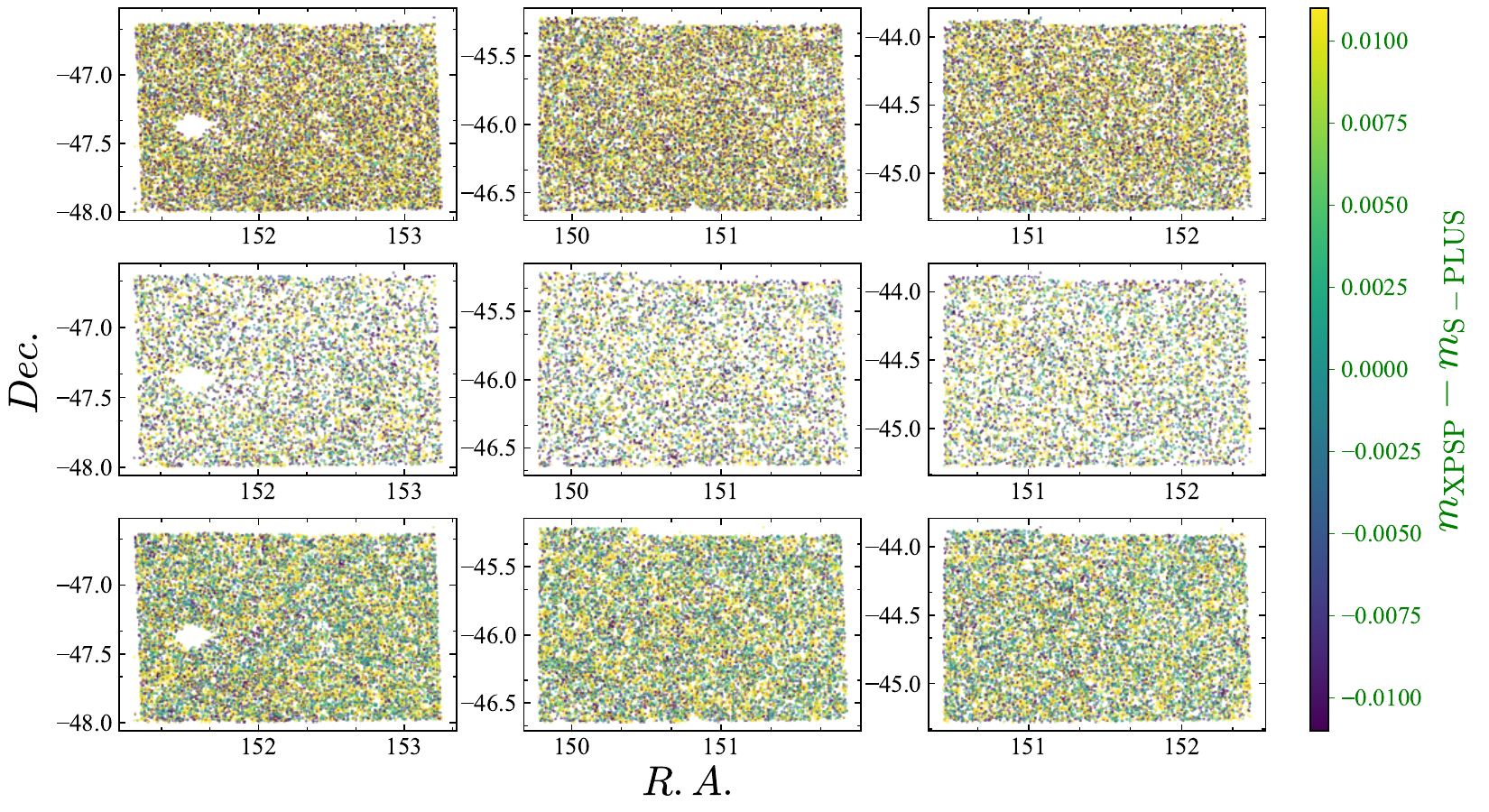}}
\caption{{\small Similar to Figure\,\ref{Fig:mxpsp_mplus}, but for the results after re-calibration.
}}
\label{Fig:mxpsp_mcorr}
\end{figure*}

\section{systematic errors in S-PLUS DR4}\label{sec:res}
In this section, we present the process of accurately measuring  systematic errors in the S-PLUS photometric data, as well as the their tracing and correction.

\subsection{Dependence on $G$ and $G_{\rm BP}-G_{\rm RP}$}
Figure\,\ref{Fig:scr} illustrates the relationship between the magnitude offsets predicted by the SCR method and the S-PLUS magnitudes, considering the $G$ magnitude and intrinsic color $(G_{\rm BP}-G_{\rm RP})_0$ of the calibration samples. We observe no discernible dependence with respect to either $G$ magnitudes or $(G_{\rm BP}-G_{\rm RP})_0$ color. This indicates that the detector possesses excellent linearity.

We also investigated the differences between the XPSP method magnitude predictions and S-PLUS magnitudes as functions of the $G$ magnitude and $G_{\rm BP}-G_{\rm RP}$ color. 
There is a slightly dependence on $G_{\rm BP}-G_{\rm RP}$ color, especially in the bluer and redder range, in the $u{\rm JAVA}$ and $g$ bands, as shown in Figure\,\ref{Fig:resi_bprp}. There is no dependence on $G$ magnitude for all the filters.
We attribute this effect to measurement errors in the response curve of the S-PLUS DR4 $u{\rm JAVA}$ and $g{\rm SDSS}$ filters, or the Gaia XP spectra themselves. Certainly, we should not overlook the influence that arises from extrapolating the XP spectra beyond the $u{\rm JAVA}$-band. Additional details regarding the influence of response curve on the XPSP method can be found in Xiao et al. (in prep.).

For the calibration of the $u{\rm JAVA}$ and $g{\rm SDSS}$ bands, we selectively choose stars from specific $G_{\rm BP}-G_{\rm RP}$ ranges of (0.5, 1.2) and (0.5, 1.8), respectively. Moreover, the fraction of stars falling outside the prescribed color range is only 2--3 per cent.

\subsection{Spatial Variations}
We plotted the spatial distribution of the difference between zero-points for the XPSP method and the S-PLUS magnitudes, as shown in Figures\,\ref{Fig:spatial_dis_pstotal_main}, \ref{Fig:spatial_dis_pstotal_mc}, \ref{Fig:spatial_dis_psf_main}, \ref{Fig:spatial_dis_psf_mc}, and \ref{Fig:spatial_dis_psf_disk}. The difference in the zero-points between the XPSP method and the S-PLUS magnitudes is computed as the median value of the difference between the XPSP predicted magnitudes and the S-PLUS magnitudes on each image. We observe strong spatial variations in the difference of the zero-points, caused by calibration errors in S-PLUS, which are more pronounced in the blue filters.
Simultaneously, we noticed spatial correlations in the differences in the zero-points between the different S-PLUS bands. The reasons for this are discussed in detail in Section\,\ref{tc}.

To quantitatively estimate calibration errors in the S-PLUS 
photometry, we consider histograms of the difference in zero-points between the XPSP method and the S-PLUS magnitudes, as shown in Figure\,\ref{Fig:hist}. By fitting a Gaussian distribution, we estimated the standard deviations for each band. Here, to better illustrate the effect, we forcibly set the overall zero-point difference to zero. During the re-calibration process, we calibrate the zero-point of the S-PLUS magnitudes to the XPSP method. These values indicate the internal precision of S-PLUS DR4, as also mentioned in \cite{2022MNRAS.511.4590A}, and listed in Table\,\ref{tab:2}.

\begin{deluxetable*}{cccccccccccc}[ht!]
\tablecaption{Internal Precision of the Photometric Calibration for the 12 S-PLUS Bands, in 
Units of mmag \label{tab:2}}
\tablehead{
\colhead{filters} & \colhead{Main$_{\rm Ap}$} & \colhead{Main$_{\rm PSF}$} & \colhead{MCs$_{\rm Ap}$} & \colhead{MCs$_{\rm PSF}$} & \colhead{disk$_{\rm PSF}$} & \colhead{Final Accuracy}}
\startdata
$g{\rm SDSS}$ & $6$  & $12$ & $10$ & $7$  & $12$  &  $0.8$ \\
$r{\rm SDSS}$ & $8$  & $16$ & $13$ & $8$  & $11$  &  $1.6$ \\
$i{\rm SDSS}$ & $6$  & $13$ & $11$ & $7$  & $13$  &  $1.0$ \\
$z{\rm SDSS}$ & $6$  & $12$ & $9$  & $7$  & $8$   &  $0.6$ \\
$u{\rm JAVA}$ & $21$ & $27$ & $21$ & $19$ & $34$  &  $6.1$ \\
$J0378$       & $23$ & $28$ & $15$ & $15$ & $32$  &  $5.7$ \\
$J0395$       & $19$ & $27$ & $20$ & $21$ & $66$  &  $4.1$ \\
$J0410$       & $14$ & $20$ & $14$ & $14$ & $12$  &  $1.9$ \\
$J0430$       & $14$ & $19$ & $10$ & $10$ & $15$  &  $2.8$ \\
$J0515$       & $9$  & $17$ & $11$ & $10$ & $14$  &  $1.1$ \\
$J0660$       & $6$  & $14$ & $13$ & $7$  & $17$  &  $2.6$ \\
$J0861$       & $6$  & $12$ & $8$  & $6$  & $9$   &  $0.9$ \\
\enddata
\end{deluxetable*}

\subsection{Tracing and Correction} \label{tc}
In order to trace the origin of the systematic errors in the S-PLUS photometry, we plot correlations between the zero-point offsets for each band pair in Figure\,\ref{Fig:r}, along with their corresponding correlation coefficients when the correlation coefficients are greater than $0.7$. We find a strong correlation between photometric bands with similar central wavelengths (e.g., $\Delta i{\rm SDSS}$ vs. $\Delta z{\rm SDSS}$); the data points are distributed closely along the one-to-one line. This phenomenon is predominantly driven by systematic errors in the reference photometric data in the respective bands. For example, the systematic errors in the S-PLUS $i$- and $z$-bands are predominantly influenced by the systematic errors in the Pan-STARRS and SDSS photometric data (e.g., the color and photometric re-calibration of SDSS Stripe 82 can be observed in \cite{2015ApJ...799..133Y,2022ApJS..259...26H} and while the photometric re-calibration of PS1 can be seen \cite{2022AJ....163..185X,2023arXiv230805774X}).

To correct the above systematic errors, we perform a smoothed interpolation algorithm with a linear kernel for each image. The magnitude correction of a certain star in the field of view is obtained by taking the magnitude offsets of the adjacent 20 calibration stars. The corrected magnitude $m^{\rm corr}$ can be computed as
\begin{eqnarray}
\begin{aligned}
    m^{\rm corr}=m^{\rm obs}+\Delta m(\rm R.A., decl.)~, \label{corr}
\end{aligned}
\end{eqnarray}
where $m^{\rm obs}$ is the observed magnitude from S-PLUS DR4, and $\Delta m(\rm R.A., decl.)$ is the position-dependent magnitude offset. 
The re-calibrated S-PLUS DR4 data is publicly available.

\section{Discussion}\label{sec:discussion}
This section applies to the S-PLUS DR4 Main Survey data, using it as an illustration for discussion.

\subsection{Final Accuracies}
Figure\,\ref{Fig:1v1} depicts a comparison of zero-points between the XPSP method and the SCR method for all twelve S-PLUS filters. The differences between these zero-points {\rm are} computed as the median value of the difference between the XPSP and SCR predicted magnitudes for each image. From inspection, all the points are consistently distributed along the one-to-one line for each band.

To quantitatively estimate the final accuracies of the re-calibration in this work, we present the difference in the zero-points between the XPSP method and SCR method, as a function of star numbers, in Figure\,\ref{Fig:precision}. Notably, the standard deviations start at higher values, then decrease and converge to stable values as the numbers of stars increase. The convergence value represents the re-calibrated accuracy using the XPSP method, which is 1--6\,mmag for each of the 12 bands. The final accuracy of the S-PLUS DR4 data in the 12 bands are similar in the Main Survey, MCs and Disk Survey, listed in last column of Table\,\ref{tab:2}.

\subsection{External Check by White Dwarf Loci}
We provide an independent check of the re-calibration using a white dwarf (WD) locus, known for its stability and uniformity at different spatial locations.

To accomplish this, we cross-match the WD catalog constructed by \cite{2022arXiv220606215G} from Gaia EDR3 with the S-PLUS DR4 catalog. We impose the criterion Galactic latitude $|b|>20^{\circ}$ for the selected WDs to ensure the best photometry is used. Specifically, we focus on the behavior of the $u{\rm JAVA}$-band, as the photometric systematics in the other bands are relatively small and difficult to examine with the WD locus. Additionally, we require that the photometric uncertainties of the $u{\rm JAVA}$-band from S-PLUS DR4 be smaller than 0.01\,mag. After the cross-match, we retain 100 WDs with good photometric quality. The stellar loci, $G-u{\rm JAVA}$ vs. $G_{\rm BP}-G_{\rm RP}$, of the selected WDs are shown in Figure\,\ref{Fig:check_wd}. From inspection, the WD locus from the S-PLUS photometry for the $u{\rm JAVA}$-band with $R.A.<90^{\circ}$ and $decl.<-40^{\circ}$ (red dots) significantly deviates from that with $R.A.>180^{\circ}$ and $decl.>-30^{\circ}$ (black dots). However, the WD loci for different positions, provided by our re-calibrated S-PLUS DR4 photometry, exhibit remarkable consistency, demonstrating the efficacy of both the XPSP method and SCR method for calibrating the photometric zero-points.

\subsection{Residuals of the Flat-field Correction} \label{flat}
S-PLUS utilizes the sky flat-fielding technique \citep{2019MNRAS.489..241M} for this correction. However, \cite{2022MNRAS.511.4590A} found spatial structures in the residuals of the flat-field correction in the Galactic $(X, Y)$ plane, and corrected them with numerical interpolation. In this study, we further investigate whether there are any systematic errors related in this plane before and after re-calibration of the S-PLUS photometry. Specifically, we focus on the $g{\rm SDSS}$, $J0515$, and $J0861$ bands as examples. We selected three images with $\texttt{ID}$ of $\texttt{iDR4\_3\_HYDRA-0161}$, $\texttt{iDR4\_3\_HYDRA-0152}$, and $\texttt{iDR4\_3\_HYDRA-0145}$ for this investigation, because they contain the highest number of reference stars, approximately 10,000 to 20,000 stars. The images can be retrieved from the S-PLUS 
cloud.\footnote{\url{https://splus.cloud/}}

From Figure\,\ref{Fig:mxpsp_mplus}, we observe distinct spatial structures in the stellar flat-fields, with variations larger than 0.01\,mag. Notably, the structures for each image differ from one another. For instance, the top panel ($\texttt{iDR4\_3\_HYDRA-0161}$) exhibits a trend of larger values in the center and smaller values at the edges. Conversely, the middle panel ($\texttt{iDR4\_3\_HYDRA-0152}$) displays a trend of smaller values in the center and smaller values at the edges. Lastly, the bottom panel ($\texttt{iDR4\_3\_HYDRA-0145}$) shows a trend of smaller values on the left side and larger values on the right side. However, despite these variations, the structures are consistent for different wavelength observations of each image. 
These structures, which may occur as residual artifacts following sky flat-field correction, can be effectively corrected during the re-calibration process, as shown in Figure\,\ref{Fig:mxpsp_mcorr}.

\section{Conclusions} \label{sec:conclusion}
In this paper, we present a re-calibration of S-PLUS photometry using millions of standards constructed by the XPSP method with corrected Gaia XP spectra. Additionally, we employ the SCR method with corrected Gaia EDR3 photometric data and spectroscopic data from LAMOST DR7 to construct a sample of about two hundred FGK dwarf standard stars per band, providing an independent validation.

During the comparison of zero-points between the XPSP method and S-PLUS photometric data, significant spatial variations of the zero-point offsets are identified, reaching up to 14--23\,mmag for the blue filters ($u{\rm JAVA}$, $J0378$, $J0395$, $J0410$ and $J0430$), 6--8\,mmag for the SDSS-like filters ($g{\rm SDSS}$, $r{\rm SDSS}$, $i{\rm SDSS}$ and $z{\rm SDSS}$), and 6--9\,mmag for the redder filters ($J0515$, $J0660$ and $J0861$).

Similarly, when comparing the zero-points between the XPSP and SCR methods, we find minor differences in zero-point offsets, approximately 3--6\,mmag for the blue filters, 1--2\,mmag for the SDSS-like filters, and 1--3\,mmag for the redder filters. These results show that the re-calibration achieves an accuracy of approximately 1 to 6\,mmag, when using the XPSP method in this work.

To validate our re-calibration results, we examine the color locus of white dwarfs, and as expected, the distribution of white dwarfs after re-calibration on the color-color diagram appears more compact than before calibration. Additionally, we discuss the minor systematic errors related to CCD position, and identify almost no remaining residuals in the flat-field correction of the S-PLUS photometry. The corrected S-PLUS DR4 photometric data will provide a solid data foundation for conducting scientific research that relies on high-calibration precision.

Overall, our results underscore the effectiveness of the XPSP method paired with the SCR method in improving calibration precision for wide-field surveys, when combined with Gaia photometry and XP spectra. The SCR method is not affected by the accuracy of the transmission curve, and can provide a more robust test and correction for magnitude- or color-dependent systematic errors presented in the photometry data. We propose that future releases of S-PLUS photometry should incorporate the XPSP method paired with the SCR method in their calibration process.
 
\begin{acknowledgments}
This work is supported by the National Natural Science Foundation of China through the project NSFC 12222301, 12173007 and 11603002,
the National Key Basic R\&D Program of China via 2019YFA0405503 and Beijing Normal University grant No. 310232102. 
We acknowledge the science research grants from the China Manned Space Project with NO. CMS-CSST-2021-A08 and CMS-CSST-2021-A09.
T.C.B. acknowledges partial support from grant PHY 14-30152; Physics Frontier Center/JINA Center for the Evolution of the Elements (JINA-CEE), and from OISE-1927130: The International Research
Network for Nuclear Astrophysics (IReNA), awarded by the US National Science Foundation. 
G.L. acknowledges FAPESP (procs. 2021/10429-0 and 2022/07301-5). F.A-F. acknowledges funding for this work from FAPESP grant 2018/20977-2.
S-PLUS as a whole acknowledges grant 2019/26492-3 from FAPESP

The S-PLUS project, including the T80-South robotic telescope
and the S-PLUS scientific survey, was founded as a partnership between
the Fundação de Amparo à Pesquisa do Estado de São Paulo
(FAPESP), the Observatório Nacional (ON), the Federal 
University of Sergipe (UFS), and the Federal University of Santa Catarina
(UFSC), with important financial and practical contributions from
other collaborating institutes in Brazil, Chile (Universidad de La
Serena), and Spain (Centro de Estudios de Física del Cosmos de
Aragón, CEFCA). We further acknowledge financial support from
the São Paulo Research Foundation (FAPESP), Fundação de Amparo
à Pesquisa do Estado do RS (FAPERGS), the Brazilian National 
Research Council (CNPq), the Coordination for the Improvement of
Higher Education Personnel (CAPES), the Carlos Chagas Filho Rio
de Janeiro State Research Foundation (FAPERJ), and the Brazilian
Innovation Agency (FINEP). Members of the S-PLUS collaboration are grateful for the contributions from CTIO staff in helping in the construction, commissioning, and maintenance
of the T80-South telescope and camera.

This work has made use of data from the European Space Agency (ESA) mission
Gaia (\url{https://www.cosmos.esa.int/gaia}), processed by the Gaia
Data Processing and Analysis Consortium (DPAC,
\url{https://www.cosmos.esa.int/web/gaia/dpac/consortium}). Funding for the DPAC
has been provided by national institutions, in particular the institutions
participating in the Gaia Multilateral Agreement.
Guoshoujing Telescope (the Large Sky Area Multi-Object Fiber Spectroscopic Telescope LAMOST) is a National Major Scientific Project built by the Chinese Academy of Sciences. Funding for the project has been provided by the National Development and Reform Commission. LAMOST is operated and managed by the National Astronomical Observatories, Chinese Academy of Sciences.

\end{acknowledgments}

\clearpage
\appendix
\setcounter{table}{0}   
\setcounter{figure}{0}
\renewcommand{\thetable}{A\arabic{table}}
\renewcommand{\thefigure}{A\arabic{figure}}

\section {Appendix}

\begin{figure*}[ht!] \centering
\resizebox{\hsize}{!}{\includegraphics{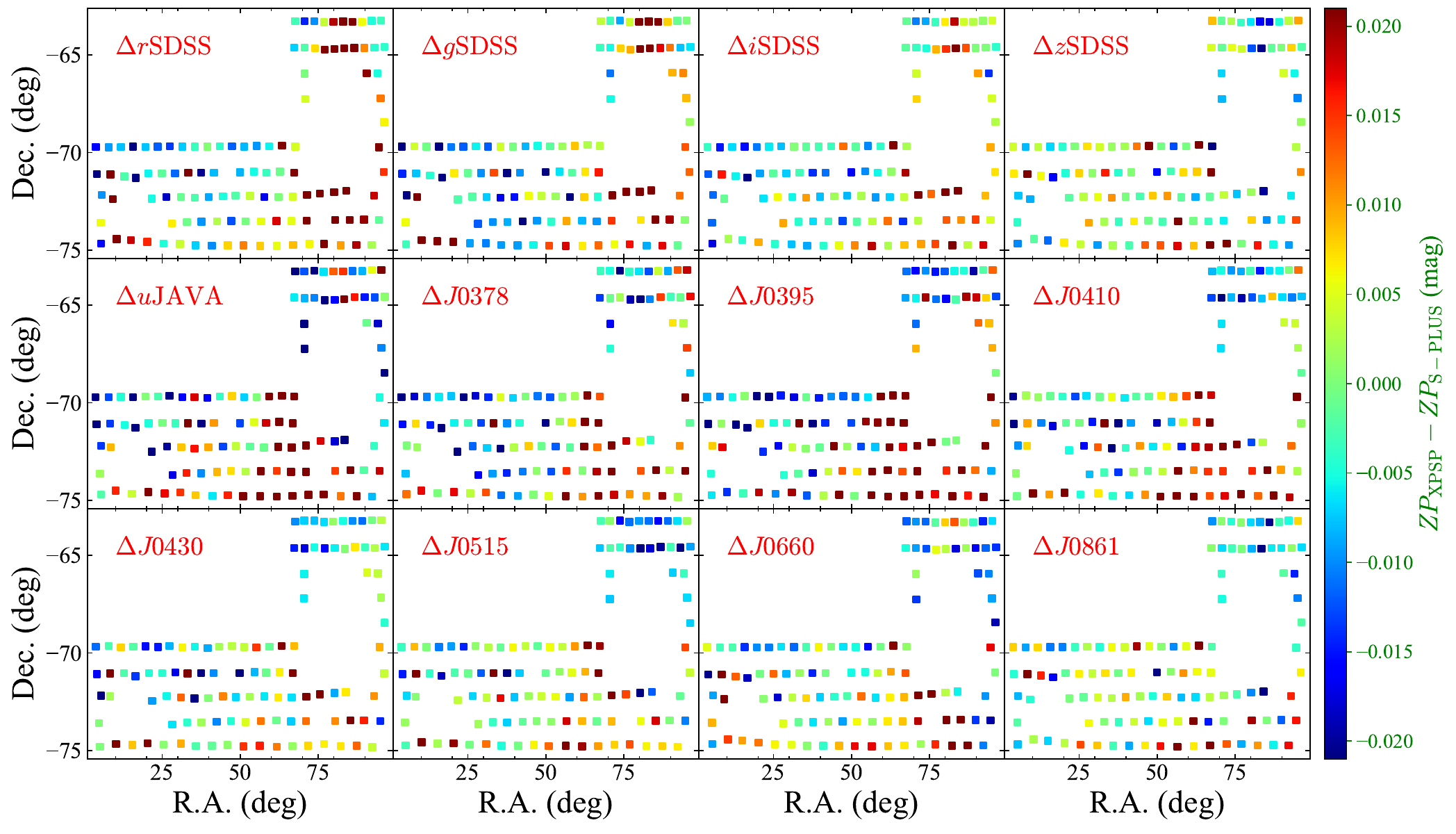}}
\caption{{\small Similar to Figure\,\ref{Fig:spatial_dis_pstotal_main}, but for the MCs with PStotal photometry.
}}
\label{Fig:spatial_dis_pstotal_mc}
\end{figure*}

\begin{figure*}[ht!] \centering
\resizebox{\hsize}{!}{\includegraphics{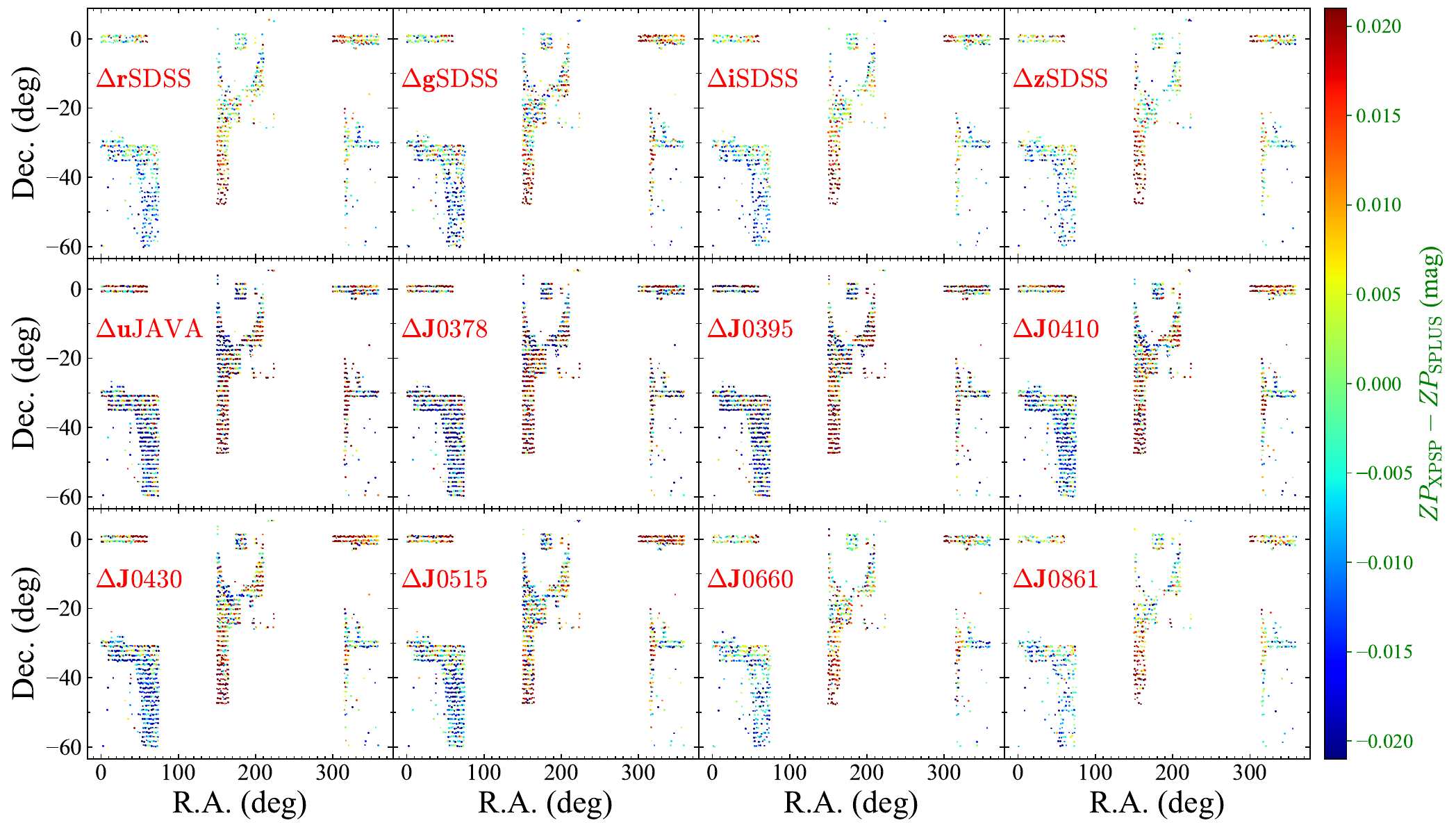}}
\caption{{\small Similar to Figure\,\ref{Fig:spatial_dis_pstotal_main}, but for the Main Survey with PSF photometry.
}}
\label{Fig:spatial_dis_psf_main}
\end{figure*}

\begin{figure*}[ht!] \centering
\resizebox{\hsize}{!}{\includegraphics{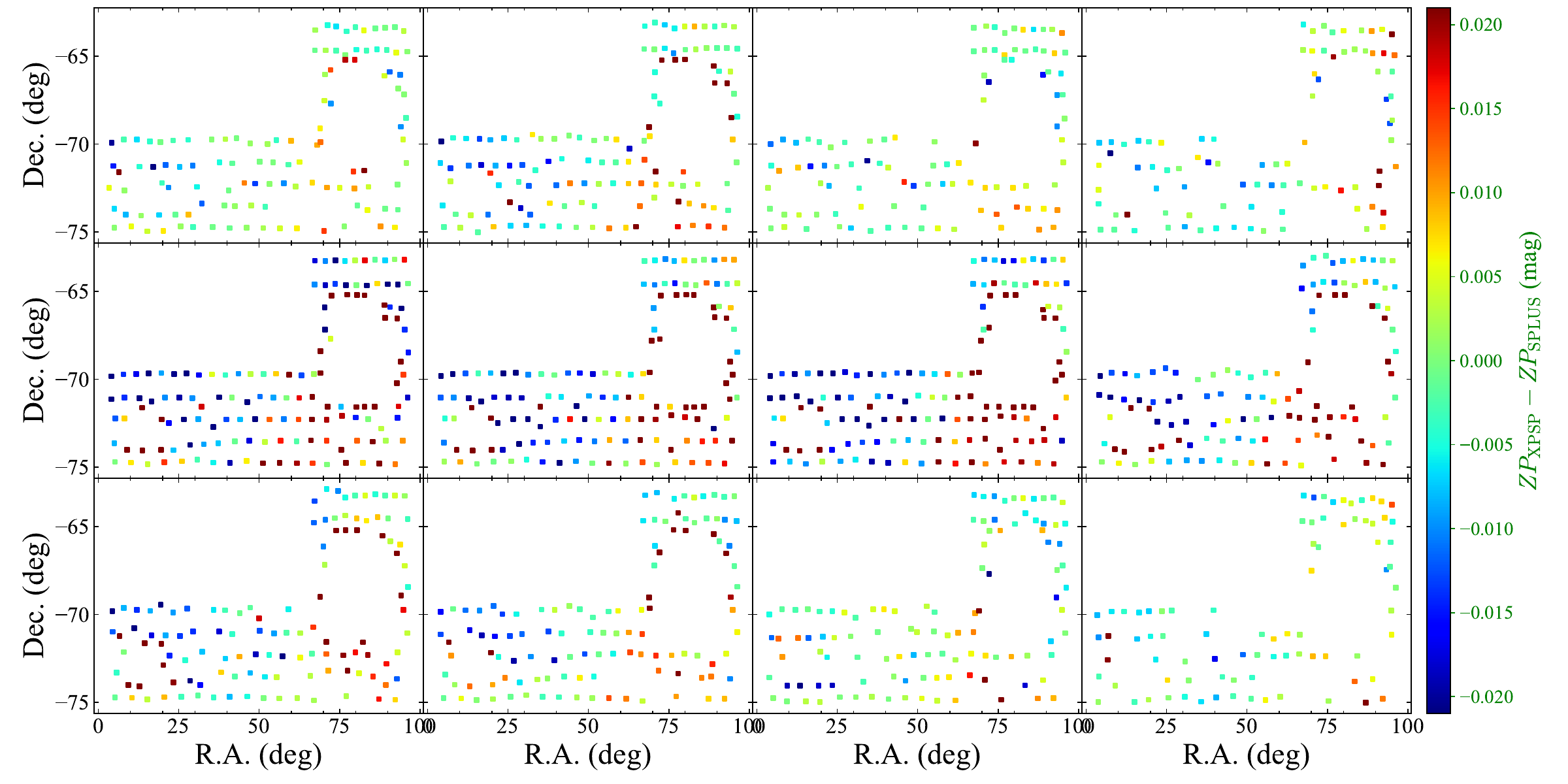}}
\caption{{\small Similar to Figure\,\ref{Fig:spatial_dis_pstotal_main}, but for the MCs fields with PSF photometry.
}}
\label{Fig:spatial_dis_psf_mc}
\end{figure*}

\begin{figure*}[ht!] \centering
\resizebox{\hsize}{!}{\includegraphics{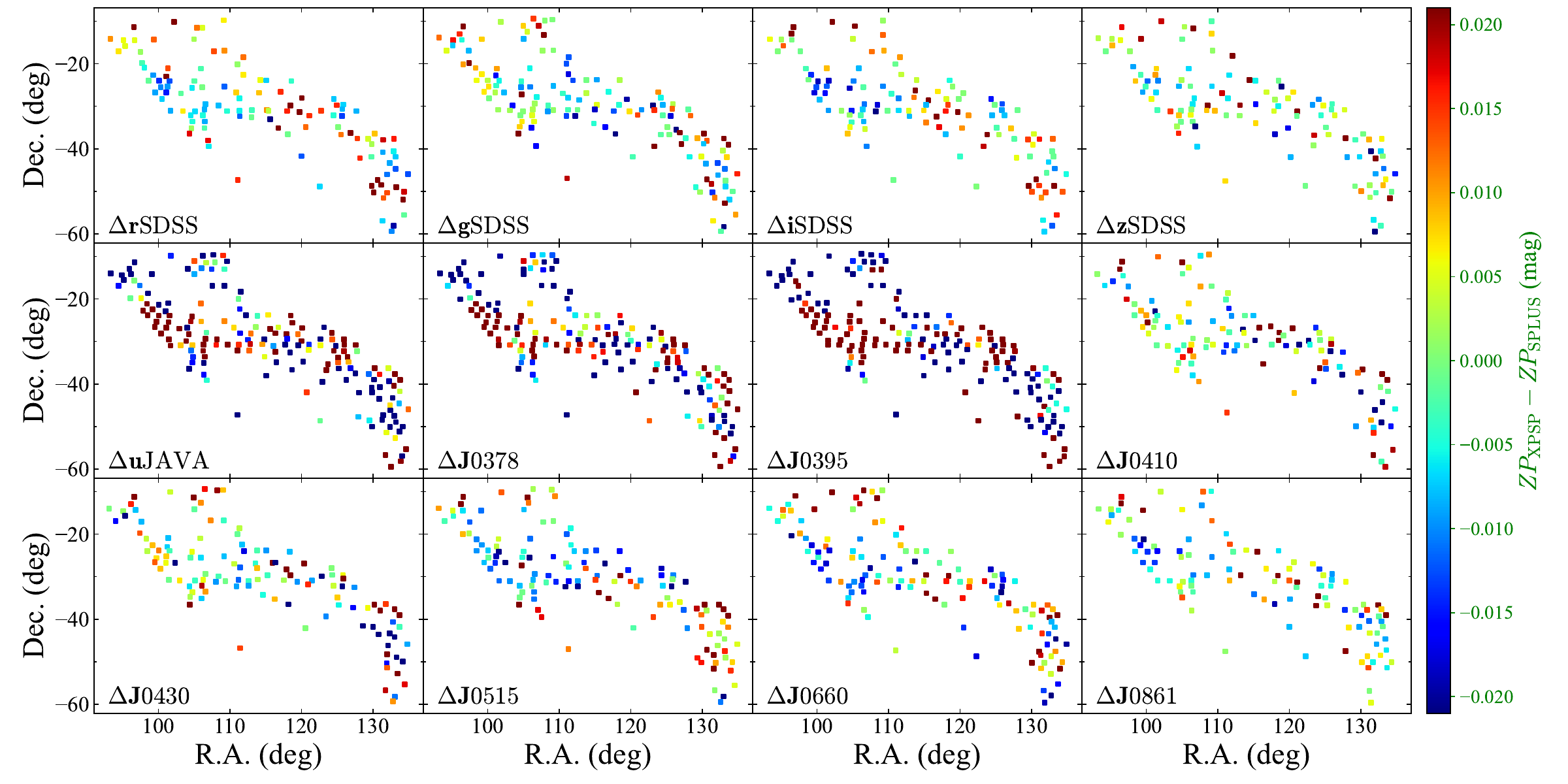}}
\caption{{\small Similar to Figure\,\ref{Fig:spatial_dis_pstotal_main}, but for the Disk Survey with PSF photometry.
}}
\label{Fig:spatial_dis_psf_disk}
\end{figure*}


\begin{thebibliography}{} \label{bib}
\bibitem[Abbott et al.(2021)]{2021ApJS..255...20A} Abbott, T.~M.~C., Adam{\'o}w, M., Aguena, M., et al.\ 2021, \apjs, 255, 20. doi:10.3847/1538-4365/ac00b3
\bibitem[Almeida-Fernandes et al.(2022)]{2022MNRAS.511.4590A} Almeida-Fernandes, F., SamPedro, L., Herpich, F.~R., et al.\ 2022, \mnras, 511, 4590. doi:10.1093/mnras/stac284
\bibitem[Bessell \& Murphy(2012)]{2012PASP..124..140B} Bessell, M. \& Murphy, S.\ 2012, \pasp, 124, 140. doi:10.1086/664083
\bibitem[Bohlin et al.(2014)]{CALSPEC14} Bohlin, R.~C., Gordon, K.~D., \& Tremblay, P.-E.\ 2014, \pasp, 126, 711. doi:10.1086/677655
\bibitem[Bohlin \& Lockwood(2022)]{CALSPEC22} Bohlin, R.~C. \& Lockwood, S.\ 2022, Instrument Science Report STIS 2022-7, 11 pages
\bibitem[Burke et al.(2018)]{2018AJ....155...41B} Burke, D.~L., Rykoff, E.~S., Allam, S., et al.\ 2018, \aj, 155, 41. doi:10.3847/1538-3881/aa9f22
\bibitem[Carrasco et al.(2021)]{2021A&A...652A..86C} Carrasco, J.~M., Weiler, M., Jordi, C., et al.\ 2021, \aap, 652, A86. doi:10.1051/0004-6361/202141249
\bibitem[Cenarro et al.(2019)]{2019A&A...622A.176C} Cenarro, A.~J., Moles, M., Crist{\'o}bal-Hornillos, D., et al.\ 2019, \aap, 622, A176. doi:10.1051/0004-6361/201833036
\bibitem[Cenarro et al.(2014)]{2014SPIE.9149E..1IC} Cenarro, A.~J., Moles, M., Mar{\'\i}n-Franch, A., et al.\ 2014, \procspie, 9149, 91491I. doi:10.1117/12.2055455
\bibitem[Clem \& Landolt(2013)]{2013AJ....146...88C} Clem, J.~L. \& Landolt, A.~U.\ 2013, \aj, 146, 88. doi:10.1088/0004-6256/146/4/88
\bibitem[Cui et al.(2012)]{2012RAA....12.1197C} Cui, X.-Q., Zhao, Y.-H., Chu, Y.-Q., et al.\ 2012, Research in Astronomy and Astrophysics, 12, 1197. doi:10.1088/1674-4527/12/9/003
\bibitem[De Angeli et al.(2022)]{2022arXiv220606143D} De Angeli, F., Weiler, M., Montegriffo, P., et al.\ 2022, arXiv:2206.06143
\bibitem[Deng et al.(2012)]{2012RAA....12..735D} Deng, L.-C., Newberg, H.~J., Liu, C., et al.\ 2012, Research in Astronomy and Astrophysics, 12, 735. doi:10.1088/1674-4527/12/7/003
\bibitem[Finkbeiner et al.(2016)]{2016ApJ...822...66F} Finkbeiner, D.~P., Schlafly, E.~F., Schlegel, D.~J., et al.\ 2016, \apj, 822, 66. doi:10.3847/0004-637X/822/2/66
\bibitem[Fukugita et al.(1996)]{1996AJ....111.1748F} Fukugita, M., Ichikawa, T., Gunn, J.~E., et al.\ 1996, \aj, 111, 1748. doi:10.1086/117915
\bibitem[Gaia Collaboration et al.(2021a)]{2021A&A...649A...1G} Gaia Collaboration, Brown, A.~G.~A., Vallenari, A., et al.\ 2021a, \aap, 649, A1. doi:10.1051/0004-6361/202039657
\bibitem[Gaia Collaboration et al.(2021b)]{2021A&A...650C...3G} Gaia Collaboration, Brown, A.~G.~A., Vallenari, A., et al.\ 2021b, \aap, 650, C3. doi:10.1051/0004-6361/202039657e
\bibitem[Gaia Collaboration et al.(2022)]{2022arXiv220606215G} Gaia Collaboration, Montegriffo, P., Bellazzini, M., et al.\ 2022, arXiv:2206.06215
\bibitem[Gaia Collaboration et al.(2022)]{2022arXiv220800211G} Gaia Collaboration, Vallenari, A., Brown, A.~G.~A., et al.\ 2022, arXiv:2208.00211. doi:10.48550/arXiv.2208.00211
\bibitem[High et al.(2009)]{2009AJ....138..110H} High, F.~W., Stubbs, C.~W., Rest, A., et al.\ 2009, \aj, 138, 110. doi:10.1088/0004-6256/138/1/110
\bibitem[Huang et al.(2021)]{2021ApJ...907...68H} Huang, Y., Yuan, H., Li, C., et al.\ 2021, \apj, 907, 68. doi:10.3847/1538-4357/abca37
\bibitem[Huang \& Yuan(2022)]{2022ApJS..259...26H} Huang, B. \& Yuan, H.\ 2022, \apjs, 259, 26. doi:10.3847/1538-4365/ac470d
{\bibitem[Huang et al. (2023)]{huang} Huang, B., Yuan, H., Xiang M.\ in prep. \label{h}}
\bibitem[Ivezi{\'c} et al.(2007)]{2007AJ....134..973I} Ivezi{\'c}, {\v{Z}}., Smith, J.~A., Miknaitis, G., et al.\ 2007, \aj, 134, 973. doi:10.1086/519976
\bibitem[Koleva \& Vazdekis(2012)]{NGSL} Koleva, M. \& Vazdekis, A.\ 2012, \aap, 538, A143. doi:10.1051/0004-6361/201118065
\bibitem[Landolt(1992)]{1992AJ....104..372L} Landolt, A.~U.\ 1992, \aj, 104, 372. doi:10.1086/116243
\bibitem[Landolt(2009)]{2009AJ....137.4186L} Landolt, A.~U.\ 2009, \aj, 137, 4186. doi:10.1088/0004-6256/137/5/4186
\bibitem[Landolt(2013)]{2013AJ....146..131L} Landolt, A.~U.\ 2013, \aj, 146, 131. doi:10.1088/0004-6256/146/5/131
\bibitem[Liu et al.(2014)]{2014IAUS..298..310L} Liu, X.-W., Yuan, H.-B., Huo, Z.-Y., et al.\ 2014, Setting the scene for Gaia and LAMOST, 298, 310. doi:10.1017/S1743921313006510
\bibitem[L{\'o}pez-Sanjuan et al.(2019)]{2019A&A...631A.119L} L{\'o}pez-Sanjuan, C., Varela, J., Crist{\'o}bal-Hornillos, D., et al.\ 2019, \aap, 631, A119. doi:10.1051/0004-6361/201936405
\bibitem[Lu \& Yuan(2023)]{lu} Lu, X.\& Yuan, H.-B., et al.\ 2023, in prep.
\bibitem[Luo et al.(2015)]{2015RAA....15.1095L} Luo, A.-L., Zhao, Y.-H., Zhao, G., et al.\ 2015, Research in Astronomy and Astrophysics, 15, 1095. doi:10.1088/1674-4527/15/8/002
\bibitem[Marin-Franch et al.(2015)]{2015IAUGA..2257381M} Marin-Franch, A., Taylor, K., Cenarro, J., et al.\ 2015, IAU General Assembly
\bibitem[Mendes de Oliveira et al.(2019)]{2019MNRAS.489..241M} Mendes de Oliveira, C., Ribeiro, T., Schoenell, W., et al.\ 2019, \mnras, 489, 241. doi:10.1093/mnras/stz1985
\bibitem[Montegriffo et al.(2022)]{2022arXiv220606205M} Montegriffo, P., De Angeli, F., Andrae, R., et al.\ 2022, arXiv:2206.06205
\bibitem[Niu et al.(2021a)]{2021ApJ...909...48N} Niu, Z., Yuan, H., \& Liu, J.\ 2021a, \apj, 909, 48. doi:10.3847/1538-4357/abdbac
\bibitem[Niu et al.(2021b)]{2021ApJ...908L..14N} Niu, Z., Yuan, H., \& Liu, J.\ 2021b, \apjl, 908, L14. doi:10.3847/2041-8213/abe1c2
\bibitem[Niu et al.(2021c)]{2021ApJ...922..211N} Niu, Z., Yuan, H., Wang, S., et al.\ 2021c, \apj, 922, 211. doi:10.3847/1538-4357/ac2573
\bibitem[Oke \& Gunn(1983)]{1983ApJ...266..713O} Oke, J.~B. \& Gunn, J.~E.\ 1983, \apj, 266, 713. doi:10.1086/160817
\bibitem[Padmanabhan et al.(2008)]{2008ApJ...674.1217P} Padmanabhan, N., Schlegel, D.~J., Finkbeiner, D.~P., et al.\ 2008, \apj, 674, 1217. doi:10.1086/524677
\bibitem[Schlegel et al.(1998)]{1998ApJ...500..525S} Schlegel, D.~J., Finkbeiner, D.~P., \& Davis, M.\ 1998, \apj, 500, 525. doi:10.1086/305772
\bibitem[Stetson(2000)]{2000PASP..112..925S} Stetson, P.~B.\ 2000, \pasp, 112, 925. doi:10.1086/316595
\bibitem[Sun et al.(2022)]{2022ApJS..260...17S} Sun, Y., Yuan, H., \& Chen, B.\ 2022, \apjs, 260, 17. doi:10.3847/1538-4365/ac642f
\bibitem[Thanjavur et al.(2021)]{2021MNRAS.505.5941T} Thanjavur, K., Ivezi{\'c}, {\v{Z}}., Allam, S.~S., et al.\ 2021, \mnras, 505, 5941. doi:10.1093/mnras/stab1452
\bibitem[Tonry et al.(2012)]{2012ApJ...750...99T} Tonry, J.~L., Stubbs, C.~W., Lykke, K.~R., et al.\ 2012, \apj, 750, 99. doi:10.1088/0004-637X/750/2/99
\bibitem[Wolf et al.(2018)]{2018PASA...35...10W} Wolf, C., Onken, C.~A., Luvaul, L.~C., et al.\ 2018, \pasa, 35, e010. doi:10.1017/pasa.2018.5
\bibitem[Wu et al.(2011)]{2011RAA....11..924W} Wu, Y., Luo, A.-L., Li, H.-N., et al.\ 2011, Research in Astronomy and Astrophysics, 11, 924. doi:10.1088/1674-4527/11/8/006
\bibitem[Xiao \& Yuan(2022)]{2022AJ....163..185X} Xiao, K. \& Yuan, H.\ 2022, \aj, 163, 185. doi:10.3847/1538-3881/ac540a
\bibitem[Xiao et al. (2023a)]{xiao} Xiao, K., Yuan, H., Huang B., et al.\ 2023a, Chinese Science Bulletin (in Chinese). doi:10.1360/TB-2023-0052
\bibitem[Xiao et al.(2023b)]{2023arXiv230805774X} Xiao, K., Yuan, H., Huang, B., et al.\ 2023b, arXiv:2308.05774. doi:10.48550/arXiv.2308.05774
\bibitem[Xu et al.(2022)]{2022ApJS..258...44X} Xu, S., Yuan, H., Niu, Z., et al.\ 2022, \apjs, 258, 44. doi:10.3847/1538-4365/ac3df6
\bibitem[Yang et al.(2021)]{2021ApJ...908L..24Y} Yang, L., Yuan, H., Zhang, R., et al.\ 2021, \apjl, 908, L24. doi:10.3847/2041-8213/abdbae
\bibitem[York et al.(2000)]{2000AJ....120.1579Y} York, D.~G., Adelman, J., Anderson, J.~E., et al.\ 2000, \aj, 120, 1579. doi:10.1086/301513
\bibitem[Yuan et al.(2013)]{2013MNRAS.430.2188Y} Yuan, H.~B., Liu, X.~W., \& Xiang, M.~S.\ 2013, \mnras, 430, 2188. doi:10.1093/mnras/stt039
\bibitem[Yuan et al.(2015a)]{2015ApJ...799..133Y} Yuan, H., Liu, X., Xiang, M., et al.\ 2015a, \apj, 799, 133. doi:10.1088/0004-637X/799/2/133
\bibitem[Yuan et al.(2015b)]{2015ApJ...799..134Y} Yuan, H., Liu, X., Xiang, M., et al.\ 2015b, \apj, 799, 134. doi:10.1088/0004-637X/799/2/134
\bibitem[Zhang \& Yuan(2020)]{2020ApJ...905L..20R} Zhang, R. \& Yuan, H.\ 2020, \apjl, 905, L20. doi:10.3847/2041-8213/abccc4
\bibitem[Zhao et al.(2012)]{2012RAA....12..723Z} Zhao, G., Zhao, Y.-H., Chu, Y.-Q., et al.\ 2012, Research in Astronomy and Astrophysics, 12, 723. doi:10.1088/1674-4527/12/7/002
\bibitem[Zheng et al.(2018)]{Zheng18} Zheng, J. and 6 colleagues.\ The SAGE photometric survey: technical description.\ Res Astron Astrophys, 2018, 18. doi:10.1088/1674-4527/18/12/147
\bibitem[Zheng et al.(2019)]{Zheng19} Zheng, J. and 6 colleagues.\ Test area of the SAGE survey.\ Res Astron Astrophys, 2019, 19. doi:10.1088/1674-4527/19/1/3
\end{thebibliography}
\end{document}